\pdfoutput=1

\RequirePackage{fix-cm}
\documentclass[a4paper]{svjour3}                     
\smartqed  
\usepackage{geometry}
\usepackage[utf8]{inputenc}
\usepackage{graphicx}
\usepackage{lineno,hyperref}
\usepackage{subfigure}
\usepackage{amsmath}
\usepackage{underscore}
\usepackage{stfloats}
\usepackage{float}
\usepackage{enumerate}
\usepackage{lineno}
\usepackage{multirow}
\usepackage[numbers,sort&compress]{natbib}
\usepackage{color}
\makeatletter
\renewcommand{\@thesubfigure}{\hskip\subfiglabelskip}
\makeatother

\begin{document}
\title{Two New Stenosis Detection Methods of Coronary
Angiograms
}


\author{Yaofang	Liu	\and Xinyue	Zhang \and Wenlong	Wan \and Shaoyu	Liu	\and Yingdi	Liu	\and Hu	Liu \and Xueying Zeng \and Qing Zhang
}


\institute{Qing Zhang\at
              Department of Cardiology, Qilu Hospital (Qingdao), Cheeloo College of Medicine, Shandong University, Qingdao, Shandong, China.\\
              \email{qingzhang2019@foxmail.com}           \\
           \and
          Yaofang	Liu	\and Xinyue	Zhang 	\and Yingdi	Liu	 \and Xueying Zeng \at School of Mathematical Sciences, Ocean University of China, Qingdao, Shandong, China
          \and 
          Wenlong	Wan \and Shaoyu	Liu \at School of Computer Science and Technology, Ocean University of China, Qingdao, Shandong, China
          \and  Hu	Liu \at School of Material Science and Engineering, Ocean University of China, Qingdao, Shandong, China
} 

\date{Received: date / Accepted: date}

\maketitle

\begin{abstract}
\noindent \textbf{Purpose}  Coronary angiography is the ``gold standard'' for diagnosing coronary artery disease (CAD). At present, the methods for detecting and evaluating coronary artery stenosis cannot satisfy the clinical needs, e.g.,  there is no prior study of detecting stenoses in prespecified vessel segments, which is necessary in clinical practice.

\noindent \textbf{Methods}  Two vascular stenosis detection methods are proposed to assist the diagnosis. The first one is an automatic method, which can automatically extract the entire coronary artery tree and mark all the possible stenoses. The second one is an interactive method. 
With this method, the user can choose any vessel segment to do further analysis of its stenoses.

\noindent \textbf{Results}  Experiments show that the proposed methods are robust for angiograms with various vessel structures. The precision, sensitivity, and $F_1$ score of the automatic stenosis detection method are 0.821, 0.757, and 0.788, respectively. Further investigation proves that the interactive method can provide a more precise outcome of stenosis detection, and our quantitative analysis is closer to reality.

\noindent \textbf{Conclusion} The proposed automatic method and interactive method are effective and can complement each other in clinical practice. The first method can be used for preliminary screening, and the second method can be used for further quantitative analysis. We believe the proposed solution is more suitable for the clinical diagnosis of CAD.
\keywords{Coronary angiograms \and Coronary artery diseases	\and	Automatic stenosis detection	\and	Interactive stenosis detection}
\end{abstract}

\section*{Introduction}
\label{intro}

Coronary angiography (CA) is the "gold standard" in clinical diagnosis of coronary artery disease (CAD)\cite{husmannCoronaryArteryMotion2007}, which is the leading cause of death worldwide\cite{roth2018global}, affecting over 120 million people \cite{virani2020heart}. At present, the methods for identifying coronary artery lesions, i.e., stenoses, and evaluating the stenotic degree are mostly based on the subjective estimation of doctors. It results in a large number of repetitive work that not only reduces the work efficiency but also causes significant subjective errors, which can lead to fault judgments \cite{jiangpingAssessmentCoronaryArtery2013}. Aimed to solve these problems,  people have proposed various computer-aided diagnosis methods. 

Most researchers focused on approaches for the automatic assessment of CAD. A generic framework of these approaches is as follows: (1) coronary artery tree extraction, (2) diameter calculation, and (3) analysis of the stenotic segment. The key stage that determines the speed and accuracy of such algorithms is based on coronary artery tree extraction. The related techniques can be broadly divided into three categories, including the tracking-based methods  \cite{makowski2002two,carrillo2007recursive,manniesing2007vessel}, e.g., Xiao et al. \cite{xiaoAutomaticVasculatureIdentification2013} introduced AGVT to extract the vessel centerlines by tracking from the selected seed points, using the local gray distribution and geometrical information,  e.g., Frangi et al. \cite{frangiMultiscaleVesselEnhancement1998} proposed a common vesselness enhancement technique, where the multiscale second-order local structure of an image (Hessian) is examined, and deep learning methods \cite{cruz2018novel,jo2018segmentation,fang2019greedy,xian2020main}, e.g., Nasr-Esfahani et al. \cite{nasr2018segmentation} designed a framework consisting of two convolutional neural networks (CNN), where the first CNN uses local and global image patches to establish an initial segmentation probability map, and the input of the second CNN includes the output of the first CNN as well as edge information from canny edge detector. The last kind is acknowledged promising as it is a powerful tool in computer vision \cite{danilovRealtimeCoronaryArtery2021} and has achieved higher efficiency and accuracy in similar tasks like vessel segmentation of retinal image \cite{fu2016deepvessel,yan2018three,oliveira2018retinal,jin2019dunet,wu2020nfn+,wang2020hard,chen2021retinal}. These approaches are very demanding on datasets because their performance depends largely on the accuracy of the dataset. However, it is time-consuming and burdensome to generate ground truth data for the angiography images, and there is no public dataset for this task \cite{jo2018segmentation}.

Another popular framework for automatically assessing CAD is using machine/deep learning methods to localize the stenoses directly. In prior work \cite{cong2019automated}, stenosis bounding boxes were annotated in images as stenosis localization labels. It introduced an end-to-end CNN-based workflow, where CNN and recurrent neural network models were employed for coronary artery view classification, candidate frame selection, and image-level stenosis classification. Ovalle-Magallanes et al. \cite{ovalle2020transfer} proposed a network-cut and fine-tuning hybrid method for stenosis detection in coronary angiograms employing a pre-trained CNN via transfer learning (VGG16, ResNet50, and Inception-v3). Aimed at confirming the feasibility of real-time coronary artery stenosis detection using deep learning methods, Danilov et al. \cite{danilovRealtimeCoronaryArtery2021} trained eight detectors based on different neural network architectures (MobileNet, ResNet-50, ResNet-101, Inception ResNet, NASNet) to detect the location of stenoses using angiography imaging series and assessed their performance. Nevertheless, these methods have the identical drawback: they can't give a quantitative result of the stenoses, e.g., the stenotic degree, which means they cannot meet the quantitative analysis needs for clinical diagnosising CAD.

Apart from the problems mentioned above, another considerable defect is that these existing methods only did global stenosis detection in the angiographic image.  However, medical staff needs to test a particular vascular segment for more detailed and accurate detection and quantitative analysis in practical application. In fact, to the best of our knowledge, there is no prior study of detecting stenoses in prespecified vessel segments.

Considering these issues, we introduce a novel solution that integrates automatic and interactive stenosis detection methods in this paper. Specifically, we first propose an automatic stenosis detection method to detect stenoses of the whole coronary artery tree. As for extracting the artery tree, we introduce an improved adaptive geometrical tracking  algorithm (IAGVT), which combines the advantages of tracking-based methods and filtering-based methods. On top of this global detection method, we further designed an interactive stenosis detection method that allows medical staff to choose any vessel segment to analyze stenoses more precisely.  It uses an energy function to track the centerline of the prespecified vessel segment. The first automatic method can be used for preliminary screening, and the interactive method can be used for further quantitative analysis. We argue that our solution is more suitable for the clinical diagnosis of CAD.

    

\section*{Methods}
\label{sec2}
This section will propose two stenosis detection methods: an automatic one and an interactive one.  Firstly, we introduce the fundamental algorithms of the two methods.

\subsection*{Preparation}
\label{Preparation}
Overall, to detect stenoses, we first segment the vessels. Then, we apply our methods of diameter measurement and stenotic degree evaluation to every part of the vessels. Finally, we identify the stenoses by comparing the stenotic degrees with a given standard. In this section, we propose the fundamental methods to do so.

\noindent \textbf{Image preprocessing.}
Following Xiao et al. \cite{xiaoAutomaticVasculatureIdentification2013}, we use a multiscale image enhancement method \cite{frangiMultiscaleVesselEnhancement1998} to highlight the vascular structure. However, it brings huge difficulty to accurately segment vessels that the complex and variable shape of coronary artery structure, noise caused by various factors, etc. Therefore, we propose an image preprocessing scheme including Rudin-Osher-Fatemi (ROF) \cite{rudinNonlinearTotalVariation1992}, unsharp masking (UM) \cite{malinUnsharpMasking1977}, and contrast limited adaptive histogram equalization (CLAHE) \cite{pizerContrastlimitedAdaptiveHistogram1990}. Specifically, ROF is used for image denoising as it has proved to be one of the most successful tools to do such work \cite{khanMeshfreeAlgorithmROF2017}.
Besides, we introduce UM, a classical tool for sharpness enhancement \cite{dengGeneralizedUnsharpMasking2011}, to further highlight the blood vessels. Moreover, we use CLAHE to enhance the color retinal image. This enhancement method is widely used in ophthalmology, such as the automatic detection of micro aneurisms and retinal blood vessel segmentation  \cite{setiawanColorRetinalImage2013}.


\noindent \textbf{Extract the vessel centerline.} 
As prior efforts \cite{xiaoAutomaticVasculatureIdentification2013} show, the adaptive geometrical tracking algorithm (AGVT) is effective in this task. Based on it, we introduce an improved adaptive geometrical tracking algorithm (IAGVT). 

To start the tracking, we need to get a seed point. Following AGVT, we first detect ridge points of the vascular structure, which are the gray maximum points perpendicular to the direction of the vessels. The gradient of the local maximum point in the image is zero, and its Hessian matrix is negative. Thus, if the point $\left( \varepsilon ,\eta \right) ( x<\varepsilon <x+1,y<\eta <y+1)  $ satisfies the following conditions:
\begin{equation}
	\begin{aligned}
		& \nabla ( x,y ) \nabla ( x+1,y+1 ) <0 \ or \\
		& \nabla ( x+1,y) \nabla ( x,y+1) <0 ,\\
		&\lambda _i\left( x+m,y+n \right) <0, \left( i=1,2,m=0,1,n=0,1 \right) ,
	\end{aligned}
\end{equation}
where $\nabla(x,y)$ is the gray gradient of point $(x,y)$ and $\lambda_i(x,y)$ are the eigenvalues of the Hessian matrix of point $(x, y)$, then $\left( \varepsilon ,\eta \right)$ can be considered as a local maximum point. The pixel $(x,y)$, as its approximate solution, is defined as a ridge point. After that, a seed point $P_0$ can be randomly chosen from ridge points.

Next, we track by identifying the tracking direction and finding the next tracking point. The initial tracking direction can be obtained from the gray information near the seed point. Concretely, take the seed point as the center and search for the gray maximum point $P^+$ on the circle with radius $d$. $P^+$ is the first point of forward tracking, the forward initial tracking direction $u^+$ and angle $\theta ^+$ can be expressed as:
\begin{equation}
	\begin{aligned}
		u\left( P^+ \right) =\frac{P^+-P}{\lVert P^+-P \rVert}=\left( \cos \theta ^+,\sin \theta ^+ \right). 
	\end{aligned}
\end{equation}

\begin{figure}[h]
	\centering
	\includegraphics[width=0.45\textwidth]{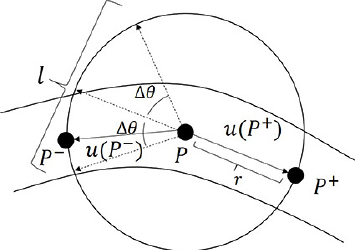}
	\caption{ Initial direction detection}
	\label{Initial direction detection}
\end{figure}
Then, we search for the local maximum point $P^-$ on arc $l(2\pi -\theta ^+-\varDelta \theta, 2\pi -\theta ^++\varDelta \theta)$  centered on the opposite direction $\left( 2\pi -\theta ^+ \right)$  of the forward tracking angle $\theta^+$.
The backward direction of the initial tracking $u^-$ can be calculated as:
\begin{equation}
	\begin{aligned}
		u\left( P^- \right) =\frac{P^--P}{\lVert P^--P \rVert}.
	\end{aligned}
\end{equation}
The search process is shown in Fig.\;\ref{Initial direction detection}.

After that, we keep track from the last point to the next point. Any tracking direction is determined by: 
\begin{equation}
	\begin{aligned}
		u_k =\frac{P_k-P_{k-1}}{\lVert P_k-P_{k-1} \rVert}.
	\end{aligned}
\end{equation}

Then, we search for the local maximum point $P_{k+1}$ on arc $l_k$($\theta_k-\Delta\theta,\theta_k+\Delta\theta$). To prevent the over-tracking beyond the vessel area in AGVT, we condition $P_{k+1}$ as follows:
\begin{equation}\label{cond1}
	\begin{aligned}
		\begin{cases}
			I\left( P_{k+1} \right) >I_0\\
			N_P\left(P_{k+1}\right)<\tau_P,
		\end{cases}
	\end{aligned}
\end{equation}
where $I(P_{k+1})$ is the gray value of $P_{k+1}$, $I_0$, $\tau _P$ are thresholds, and $N_P(P_{k+1})$ is the number of tracking points around $P_{k+1}$. The first condition in Eq.\eqref{cond1} is to prevent the over-tracking beyond the vessel area, and the second condition can avoid repeatedly tracking the vessel and being trapped in an endless local loop.

If only rely on gray information to orient, a few tracking points may devi    ate from the vascular center as the gray distribution is not uniform in some parts. Different from the centerline adjustment method in \cite{xiaoAutomaticVasculatureIdentification2013}, which still relies on the local gray distribution, we incorporate it  with the C-V model \cite{chanActiveContoursEdges2001}. The C-V model is the most popular region-based active contour model \cite{tianVesselActiveContour2014a}, and it identifies vessel and background regions using global region statistical information.

\begin{figure}[h]
	\centering
	\includegraphics[width=0.45\textwidth]{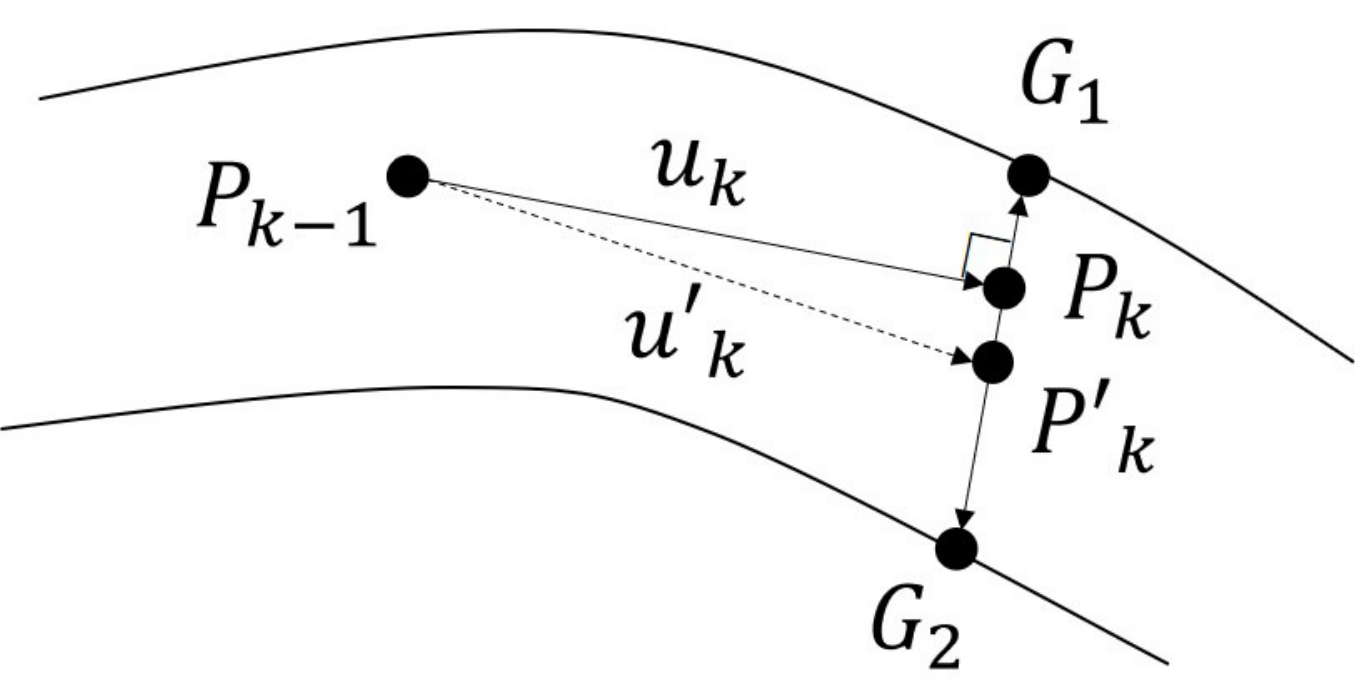}
	\caption{Centerline adjustment}
	\label{Centerline adjustment}
\end{figure}

After obtaining the vessel contour with the C-V model, the specific adjustment steps are: get the normal line of the vessel through the vertical direction of the current tracking direction, find the intersection points $G_1,G_2$ of the normal line and the vessel contour, then tracking point $P_{k}$ can be adjusted to:
\begin{equation}
	P_k'=\frac{G_1+G_2}{2},
\end{equation} 
meanwhile, change the tracking direction $u_{k}$ to:
\begin{equation}
	u_k'=\frac{P_k'-P_{k-1}}{\lVert P_k'-P_{k-1} \rVert}.
\end{equation}
The adjustment process is illustrated in Fig.\;\ref{Centerline adjustment}.

\begin{figure}[h]
	\centering
	\includegraphics[width=0.45\textwidth]{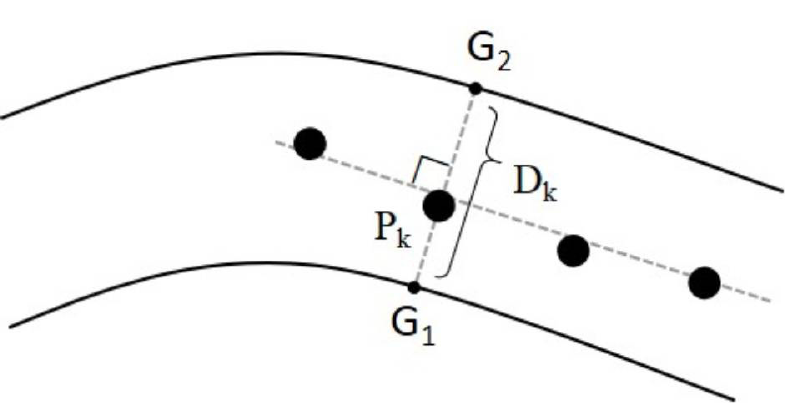}
	\caption{ Vascular diameter measurement}
	\label{Vascular diameter measurement}
\end{figure}

\noindent \textbf{Diameter measurement and stenotic degree evaluation.}
We have known how to extract the vascular skeleton and contour. Based on this, we can detect stenoses. 

In the previous section, we adjusted the tracking point to the centerline. The diameter of the target vessel tracking point can be measured according to the adjusted vascular skeleton information and the vascular contour extracted. The algorithm flow is as follows
\begin{enumerate}[(1)]
	\item Give any point $P_k$ .
	\item Calculate distances from each tracking point  $P_i (i=1,2,\ldots , n)$ on the vascular centerline to $ P_k: d_1,d_2, $ $...,  d_n$.
	\item Sort these distances and select four points with the smallest distances (if $n<4$, select $n$ points).
	\item Fit a line $l_1$  that passes through the nearest four/$n$ points using  the  least  square  method.
	\item Obtain the normal line $l_2$ of $l_1$ at $P_k$.
	\item Get the intersection points ($l_2$ with the vessel contour) $I_1$ and $I_2$.
	\item Calculate the distance $D_k$ between $I_1$ and $I_2$. $D_k$ is the diameter at $P_k$.
	
\end{enumerate}
The measurement schematic diagram is shown in Fig.\;\ref{Vascular diameter measurement}. 

The stenotic degree of the point $P_i$ is evaluated as:
\begin{equation}
	S_i=\frac{D_i}{\overline{D}},
	\label{degree}
\end{equation} 
where $\overline{D}$ is the average diameter of the corresponding vessel segment.

To assess the stenosis, we set a discriminant function:
	\begin{equation}
		\begin{aligned}
			\delta(P_i) =\begin{cases}
				1,\ \ \ \ if\ S_i<\tau_3,\\
				0,\ \ \ \ otherwise.\\
			\end{cases}
		\end{aligned},
			\label{judge}
	\end{equation}
	where $\tau_3$ is a threshold. When $\delta(P_i)$ equals zero, the point $P_i$ is viewed as stenosis. Following Neubauer et al. \cite{neubauer2010clinical}, $P_i$ is identified as stenosis when any substantial and visually evident ($>20\%$) reduction in vessel diameter. Thus, we set 0.8 as the default value for $\tau_3$ in our later experiments.

\subsection*{Automatic stenosis detection}
\label{sec3.2}
As mentioned above, firstly, a seed point is randomly chosen from the ridge points. Then, we can get the tracking point sequence of the blood vessels using IAGVT. However, our tracking method can't track the whole vessel structure yet because of the bifurcations. So, we need to add a bifurcation detection process.

\begin{figure}[h]
	\centering
	\includegraphics[width=0.45\textwidth]{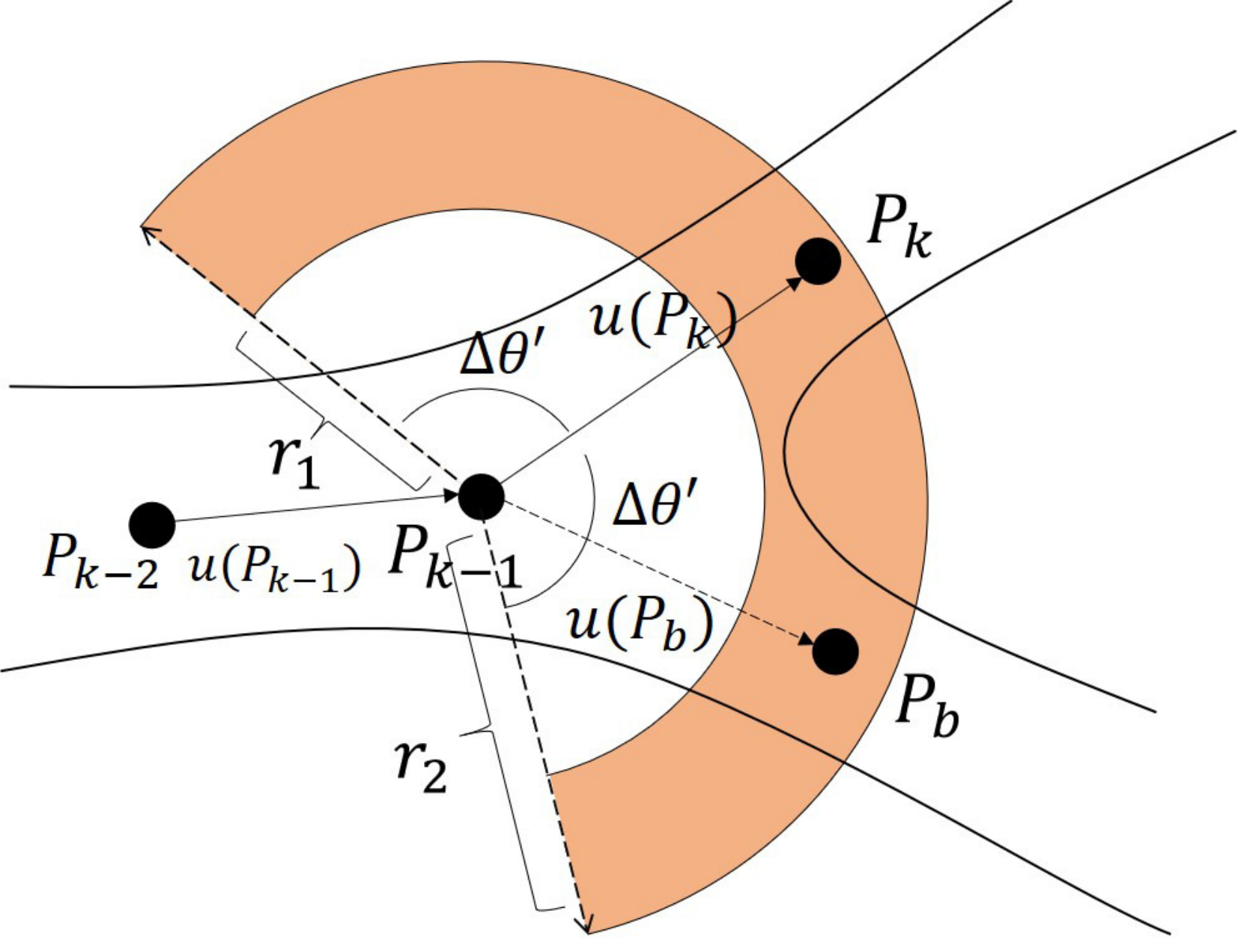}
	\caption{ Vascular bifurcation detection}
	\label{Vascular bifurcation detection}
\end{figure}

 This process includes two main steps: first, we obtain one branch point (tracking point) $P_k$ by the tracking method. Then, we need to find the other branch point. Different from Xiao et al.\cite{xiaoAutomaticVasculatureIdentification2013}, which just find another $P_k$ using the same conditions, we search in the fan ring area between angle ($\theta _k-\varDelta \theta'$,$\theta _k+\varDelta \theta'$) and radius $(r_1,r_2)$ to find a ridge point that satisfies the following conditions:
\begin{equation}\label{eq.9}
	\begin{aligned}
		\begin{cases}
			\left| \theta _b-\theta _k \right|>\tau _1\\
			\left| \theta _b-\theta _{k-1} \right|>\tau _2\\
			\lVert P_b-P_k \rVert >d\\
			N_B\left( P_b \right) <\tau _B,
		\end{cases}
	\end{aligned}
\end{equation}
where $P_b$ and $u(P_b)$  represent the detected ridge point of the new branch (the branch point) and direction of it, respectively, $N_B\left( P_b \right)$ is the number of bifurcations around the branch point $P_b$. The first three conditions in Eq.\eqref{eq.9} mean that the two branches should differ in several aspects. The last condition indicates that $N_B\left( P_b \right)$ should be smaller than threshold $\tau _B$ to avoid duplication with existing tracking. The bifurcation detection process  is shown in Fig.\;\ref{Vascular bifurcation detection}.

Now, we get the tracking point sequence of the whole vessel structure. Due to the diameters of blood vessels varying wildly, we need to segment the vessels before detecting stenoses. 

\begin{figure}[h]
	\centering
	\includegraphics{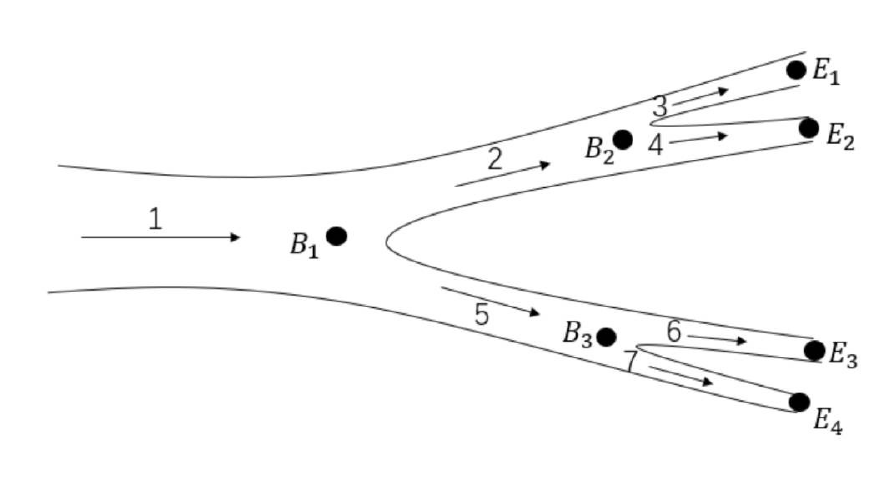}
	\caption{Tracking process. $B$ represents the bifurcation point, $E$ represents the termination point of piecewise tracking, the arrow indicates the tracking direction, and the number indicates the order of tracking.}
	\label{Vessels segmentation}
\end{figure}

In the tracking process, the tracking points on each section are continuously tracked from start to end in a geometry sense, so the tracking point sequence of a particular segment is continuous in the whole sequence.  Using this feature, we propose a concise and accurate vessel segmenting method. Specifically, we obtain vessel segments by  extracting point sequences between every two adjacent cutoff points, including bifurcation points and termination points of the tracking point sequence. A typical tracking process is shown in Fig.\;\ref{Vessels segmentation}.

\begin{figure*}[t]
	\centering
	\begin{minipage}[t]{0.25\textwidth}
		\subfigure[(a1)] { \label{fig6(a1)}     \includegraphics[width=1.3in]{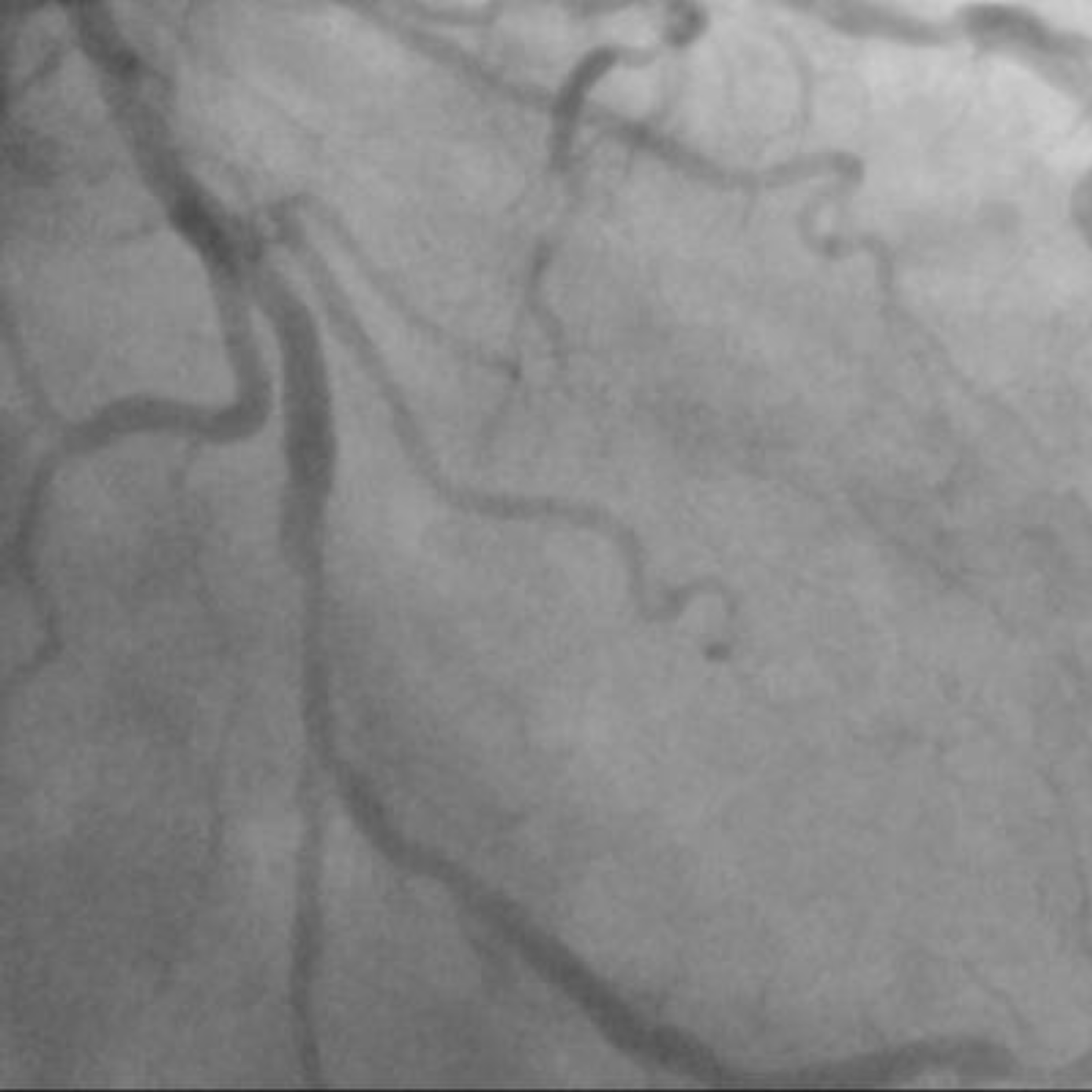} }
	\end{minipage}
	\begin{minipage}[t]{0.25\textwidth} 
		\subfigure[(a2)] { \label{fig6(a2)}     \includegraphics[width=1.3in]{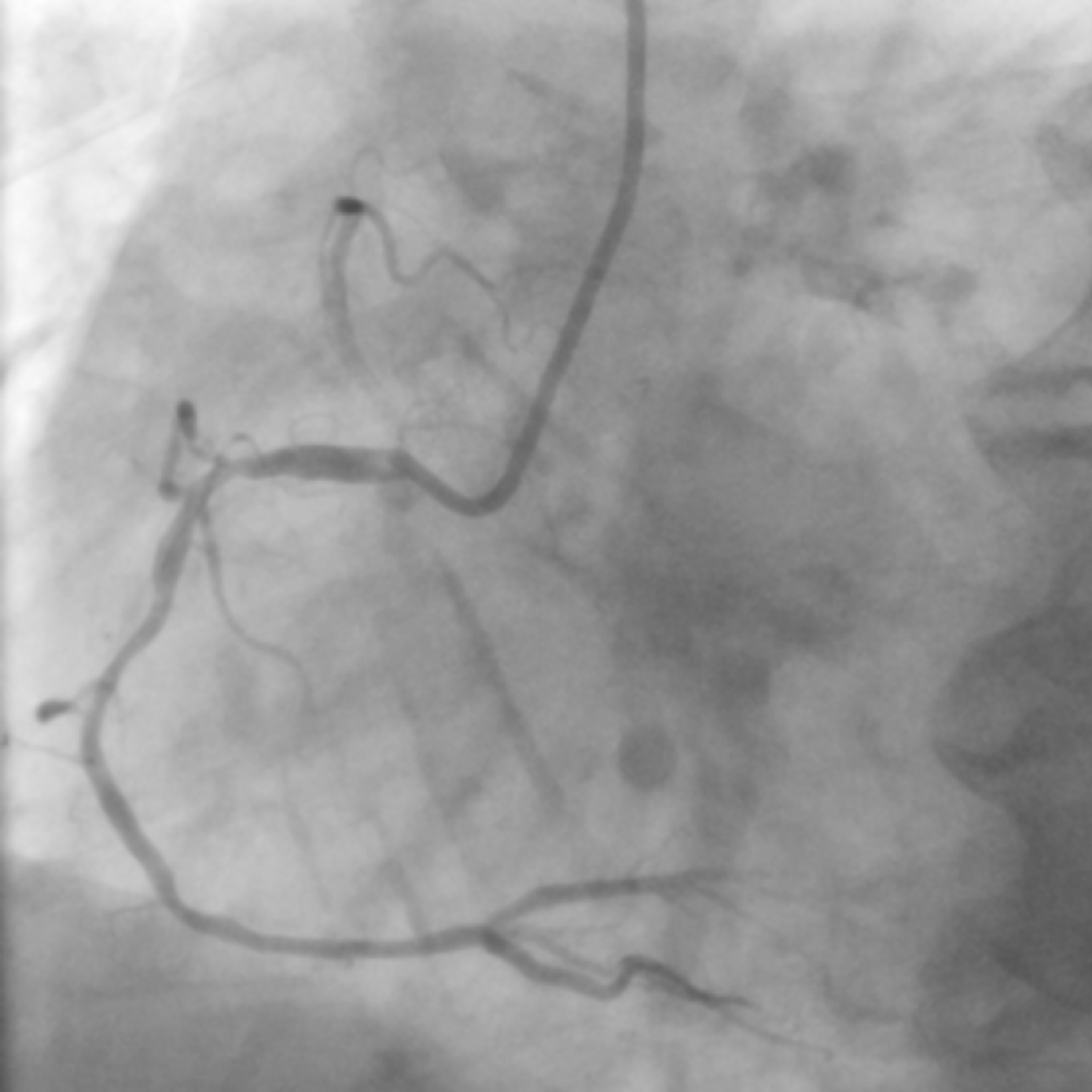} }   
	\end{minipage} 
	\begin{minipage}[t]{0.25\textwidth} 
		\subfigure[(a3)] { \label{fig6(a3)}     \includegraphics[width=1.3in]{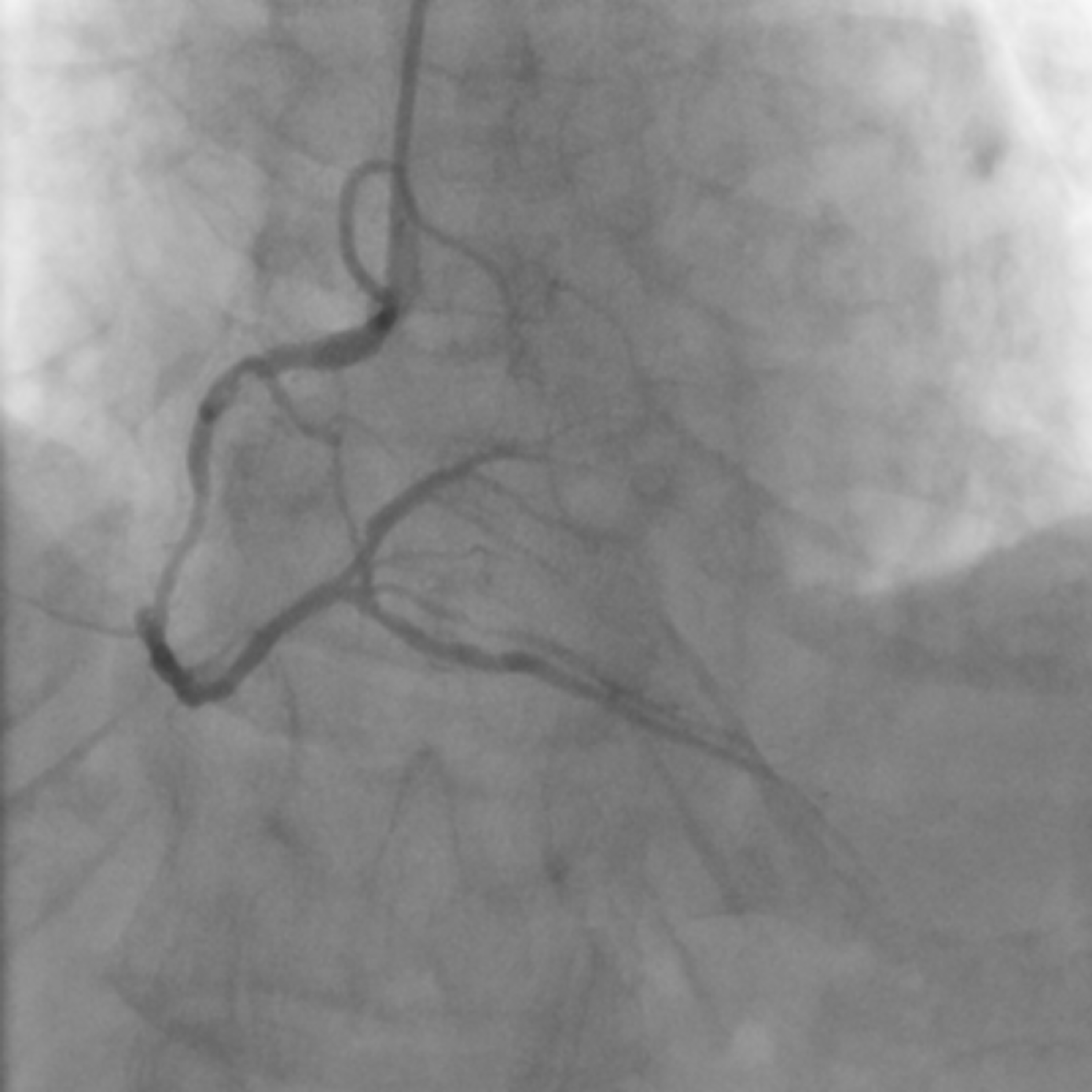} }   
	\end{minipage}
	
	\begin{minipage}{0.15\textwidth}
		\subfigure[(b1)] { \label{fig6(b1)}     \includegraphics[width=0.8in]{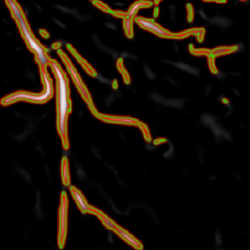} }
	\end{minipage}
	\begin{minipage}{0.15\textwidth} 
		\subfigure[(b2)] { \label{fig6(b2)}     \includegraphics[width=0.8in]{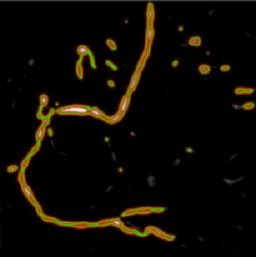} }   
	\end{minipage} 
	\begin{minipage}{0.15\textwidth} 
		\subfigure[(b3)] { \label{fig6(b3)}     \includegraphics[width=0.8in]{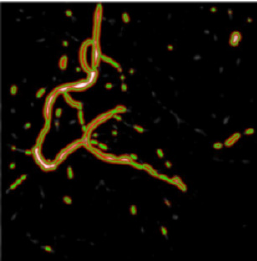} }   
	\end{minipage}
	\begin{minipage}{0.15\textwidth}
		\subfigure[(c1)] { \label{fig6(c1)}     \includegraphics[width=0.8in]{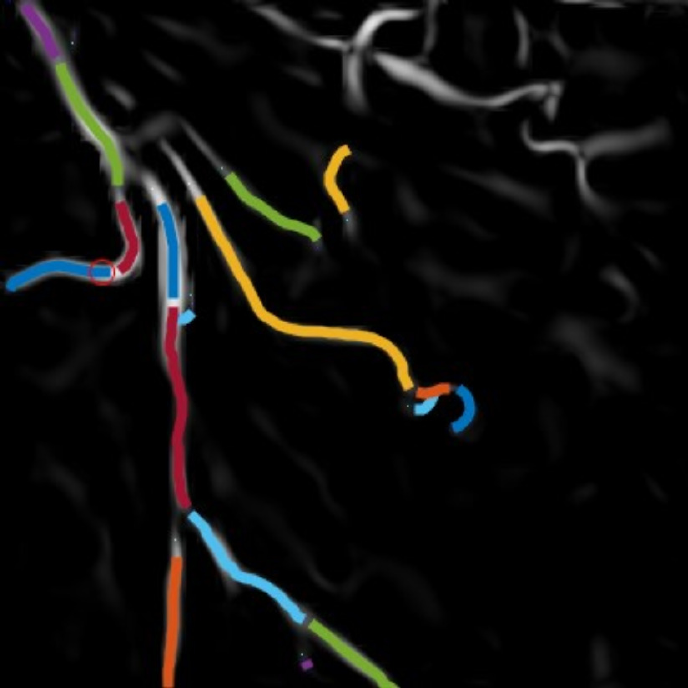} }
	\end{minipage}
	\begin{minipage}{0.15\textwidth} 
		\subfigure[(c2)] { \label{fig6(c2)}     \includegraphics[width=0.8in]{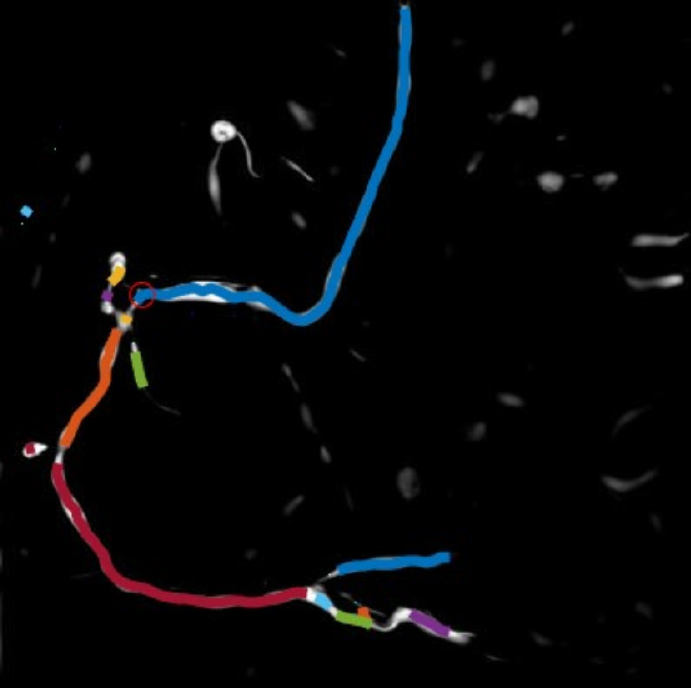} }   
	\end{minipage} 
	\begin{minipage}{0.15\textwidth} 
		\subfigure[(c3)] { \label{fig6(c3)}     \includegraphics[width=0.8in]{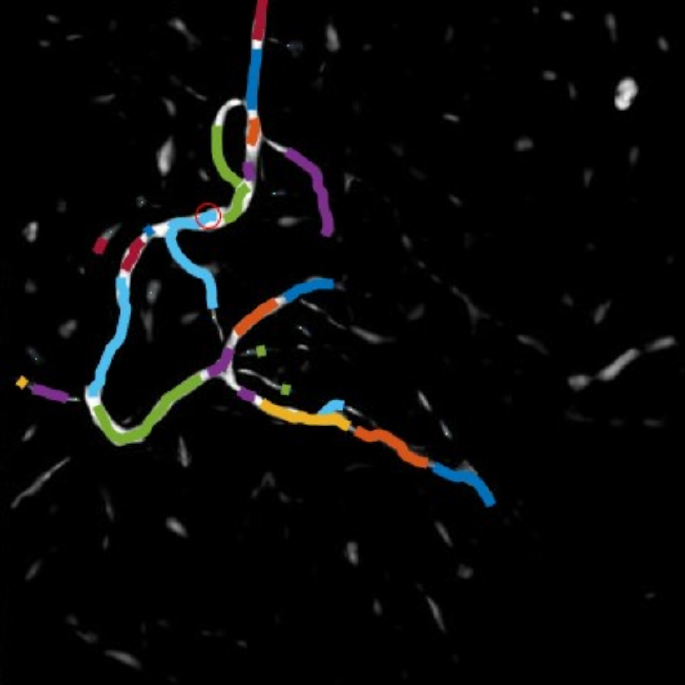} }   
	\end{minipage}
	
	\begin{minipage}{0.15\textwidth}
		\subfigure[(d1)] { \label{fig6(d1)}     \includegraphics[width=0.8in]{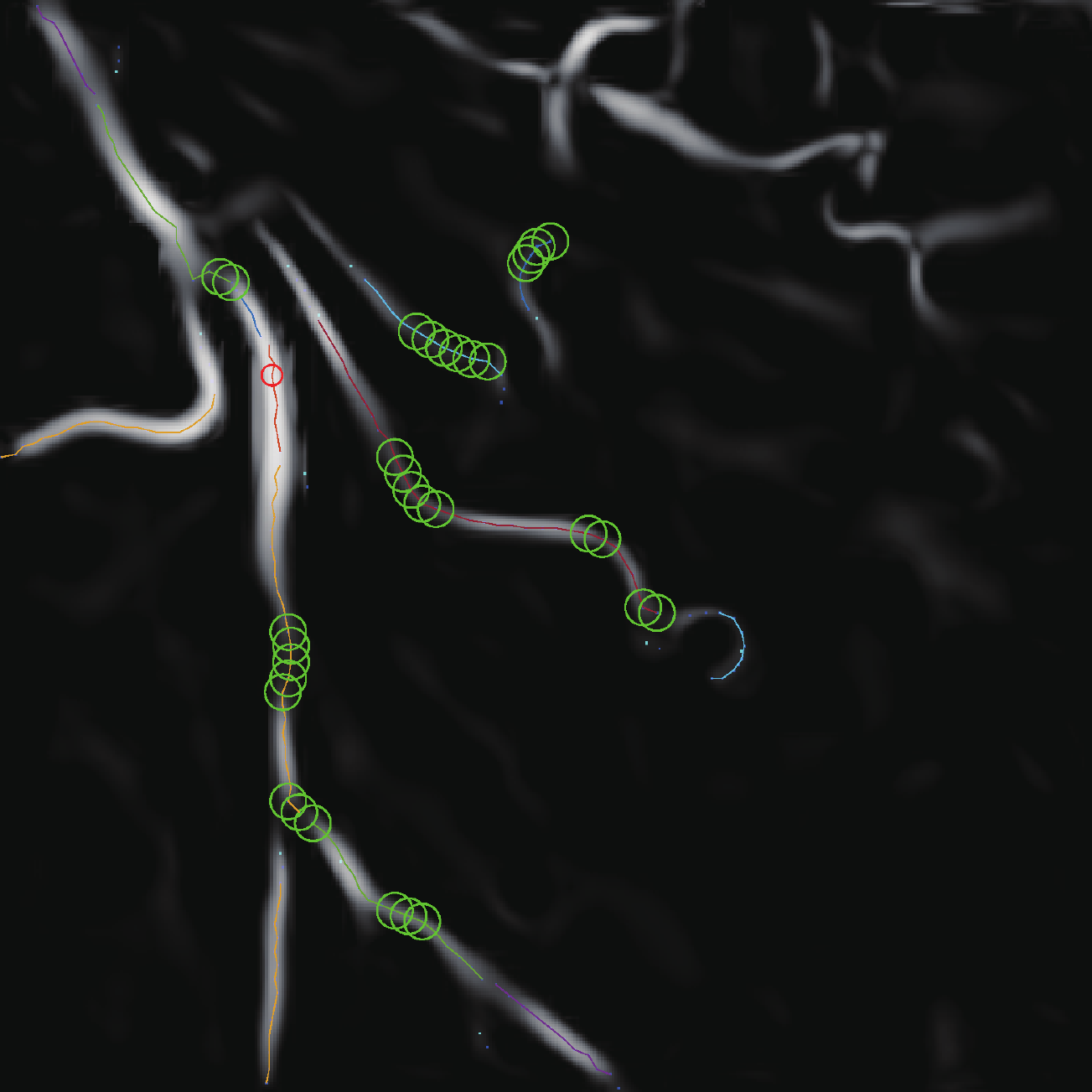} }
	\end{minipage}
	\begin{minipage}{0.15\textwidth} 
		\subfigure[(d2)] { \label{fig6(d2)}     \includegraphics[width=0.8in]{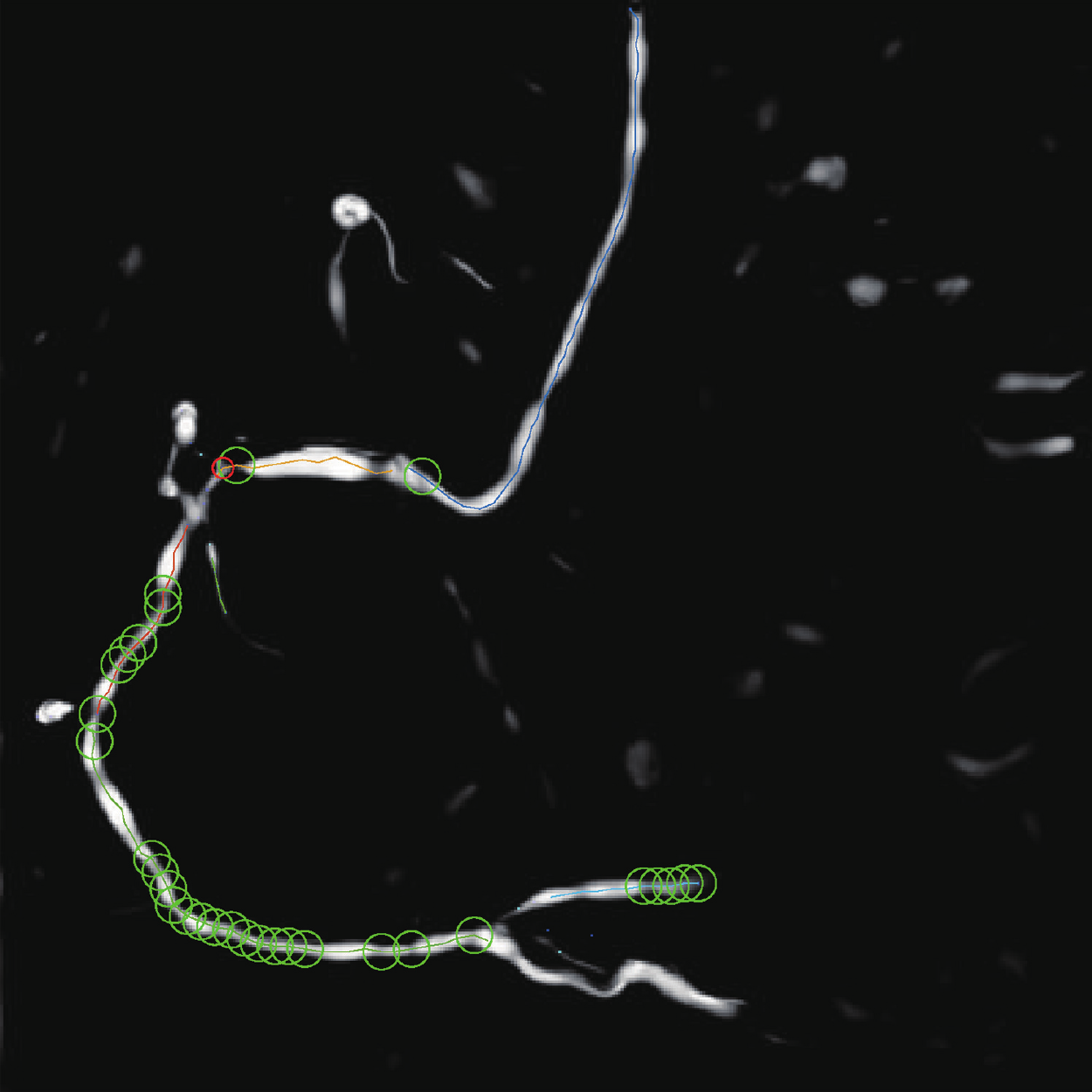} }   
	\end{minipage} 
	\begin{minipage}{0.15\textwidth} 
		\subfigure[(d3)] { \label{fig6(d3)}     \includegraphics[width=0.8in]{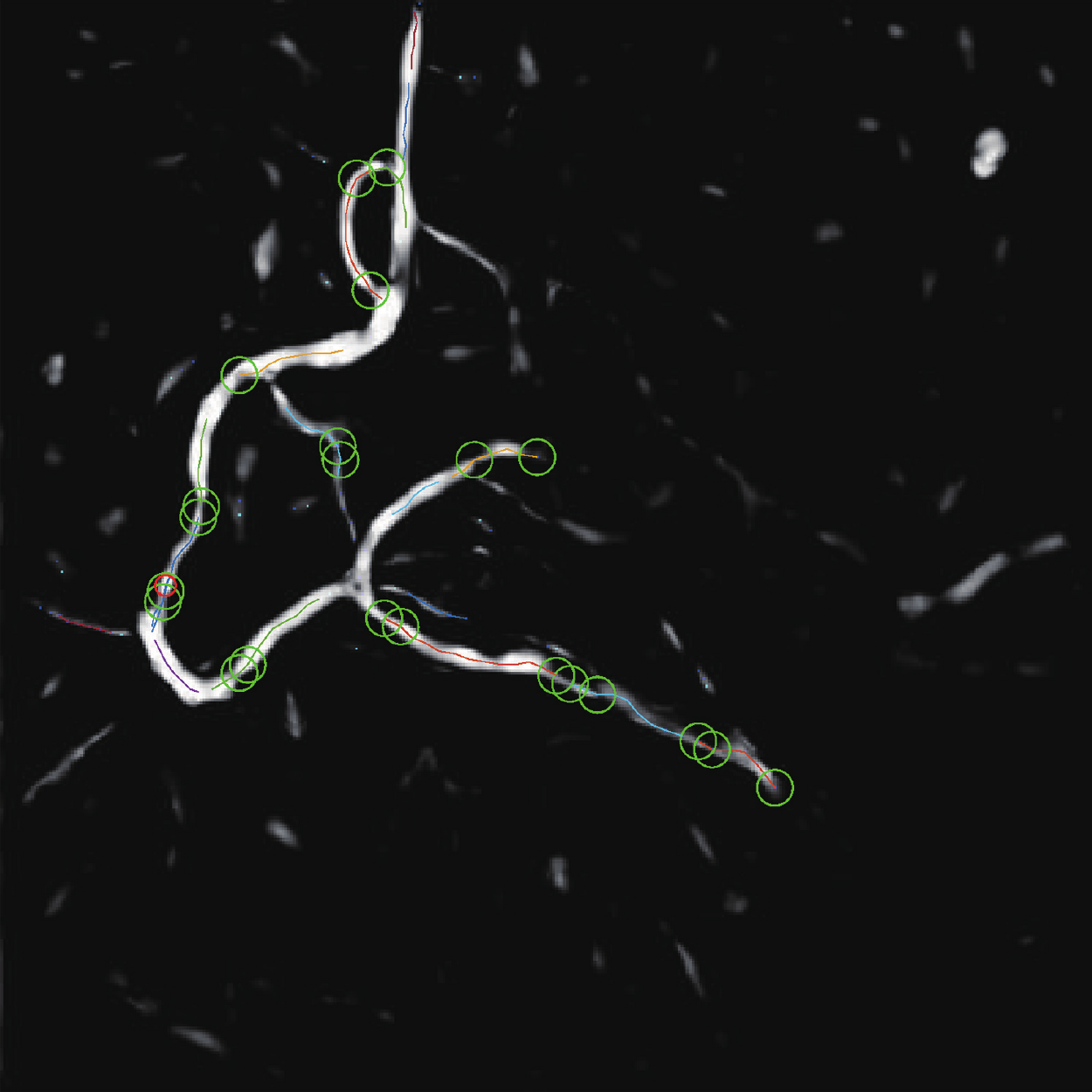} }   
	\end{minipage}
	\begin{minipage}{0.15\textwidth}
		\subfigure[(e1)] { \label{fig6(e1)}     \includegraphics[width=0.8in]{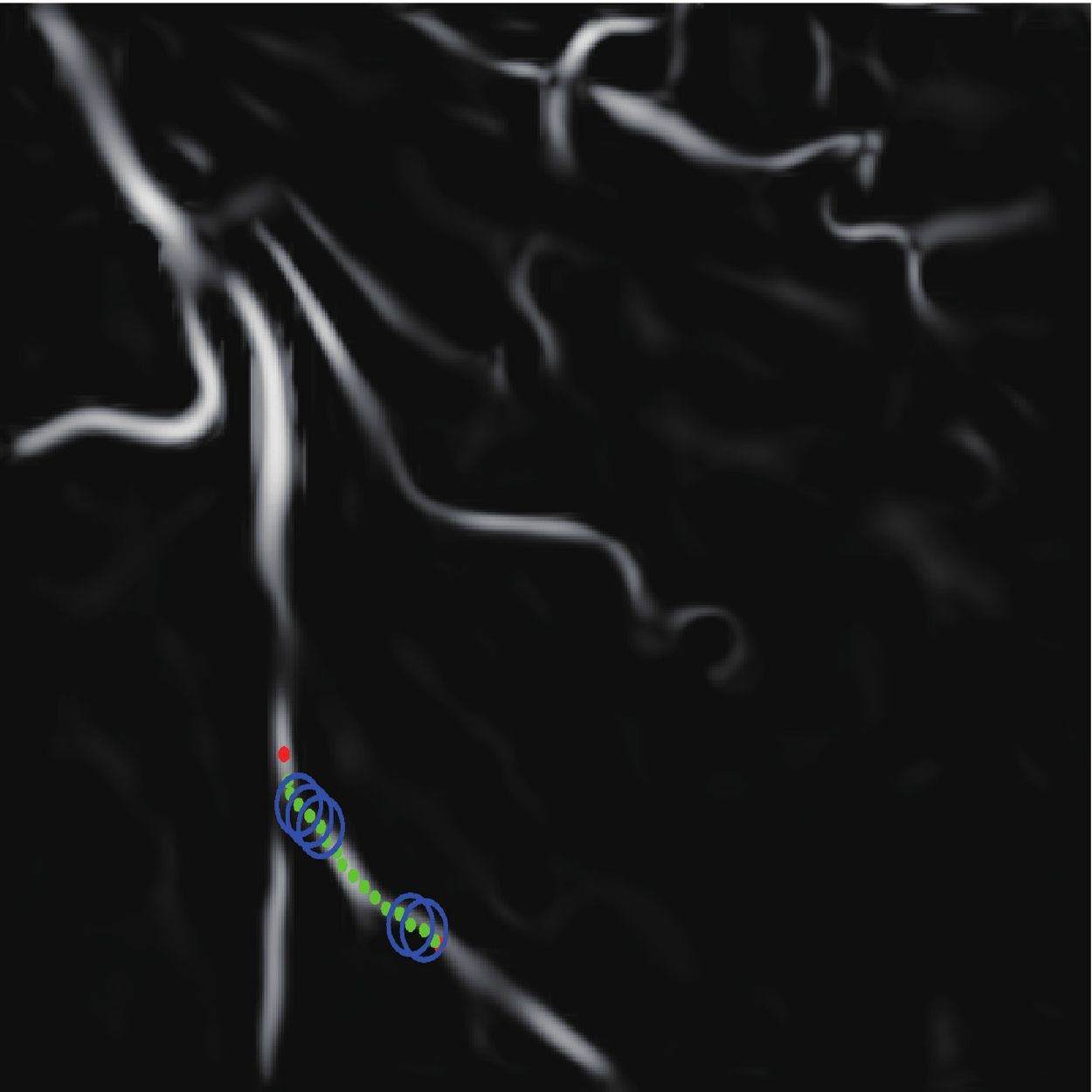} }
	\end{minipage}
	\begin{minipage}{0.15\textwidth} 
		\subfigure[(e2)] { \label{fig6(e2)}     \includegraphics[width=0.8in]{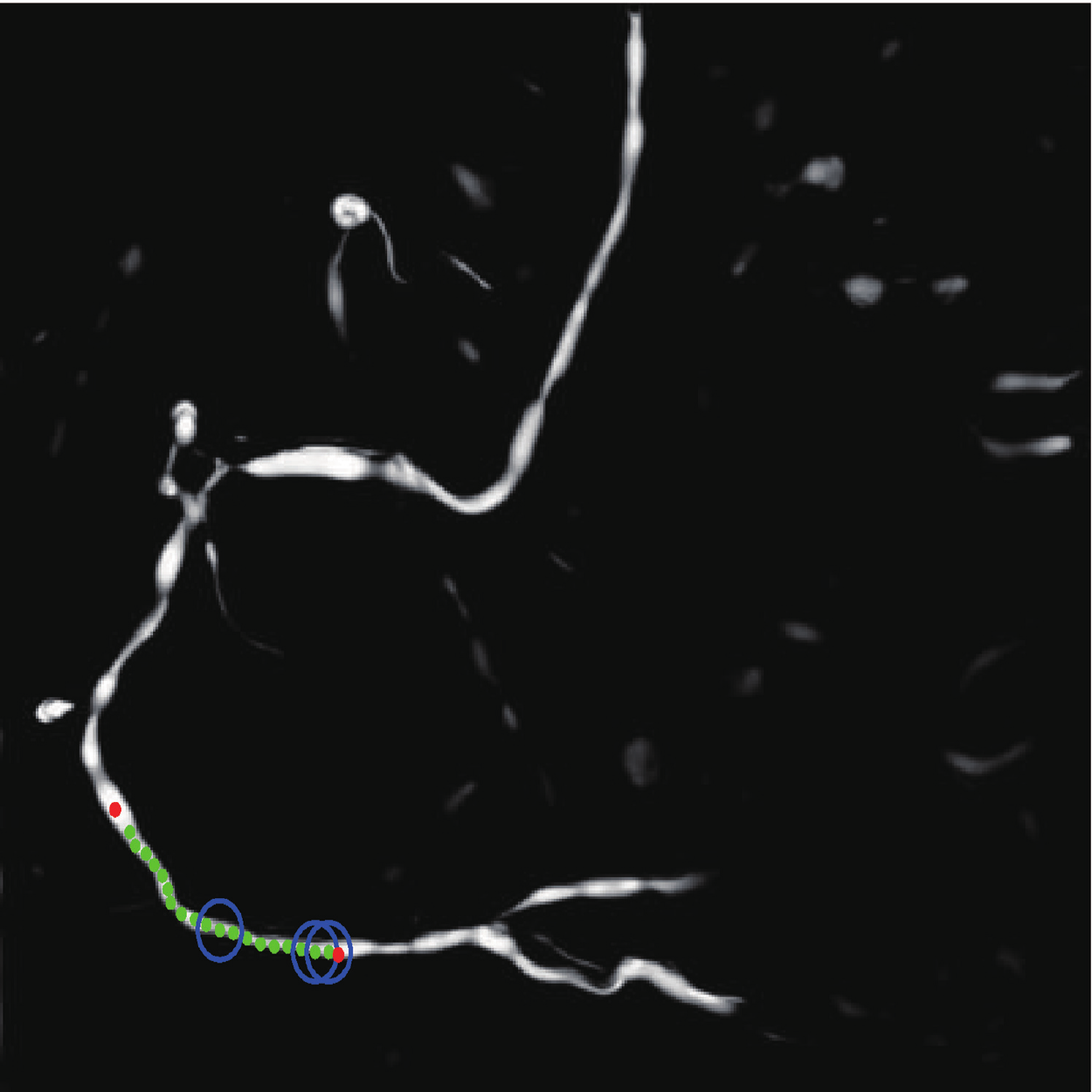} }   
	\end{minipage} 
	\begin{minipage}{0.15\textwidth} 
		\subfigure[(e3)] { \label{fig6(e3)}     \includegraphics[width=0.8in]{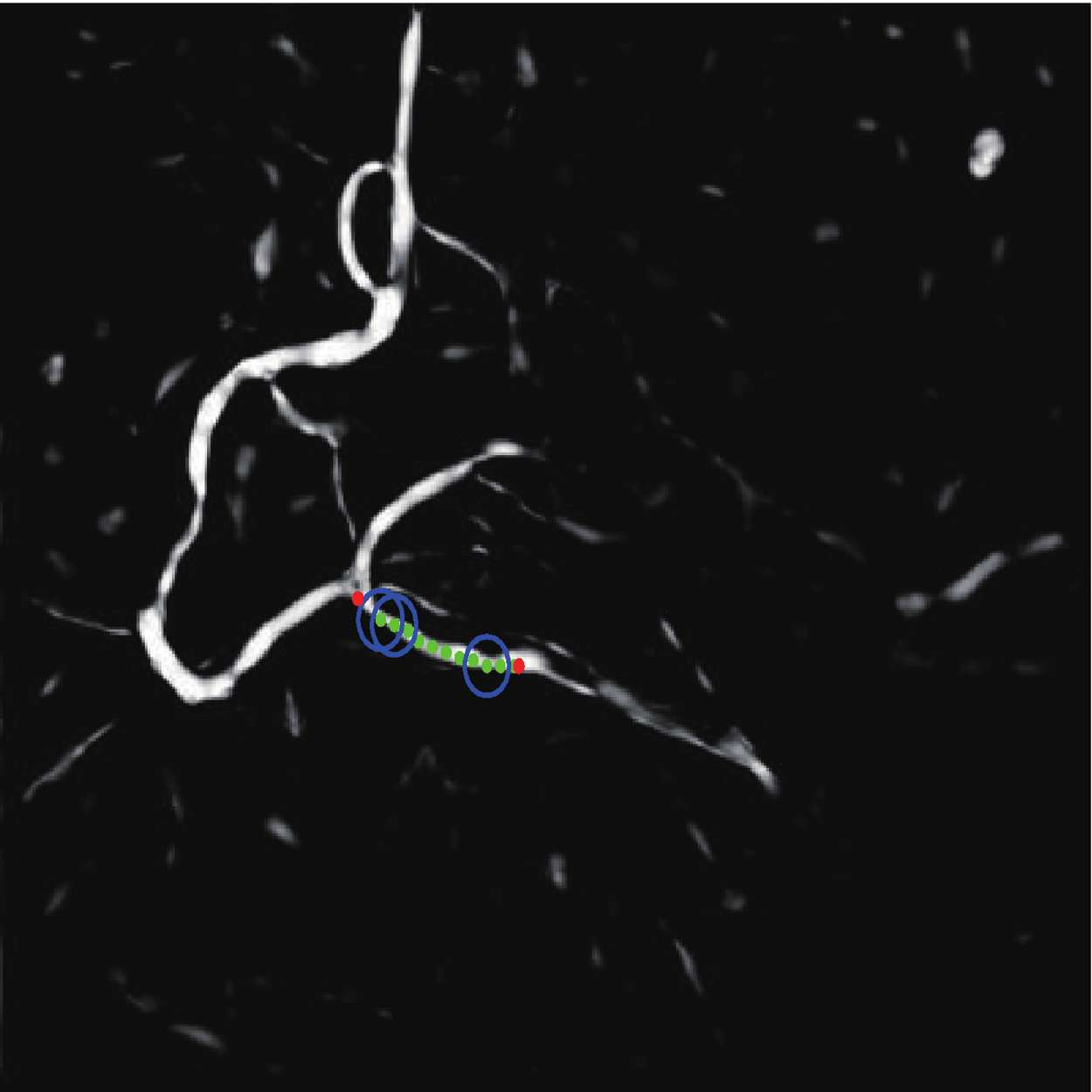} } 
	\end{minipage} 
	\centering
	\caption{Overall experimental results of our proposed solution. (a) Original images. (b) Vessel contour extraction after image preprocessing. (c) Vessel segmenting. (d) Automatic stenosis detection. (e) Interactive stenosis detection }
	\label{Methods results}
\end{figure*}

Once the segments are obtained, we can detect the stenoses by using the method of diameter measurement and stenotic degree evaluation on each vessel segment.

\begin{figure*}[h]
	\centering
	\begin{minipage}[t]{0.22\textwidth}
	    \subfigure[(a1)] {\label{fig7(a1)}     \includegraphics[width=1.2in]{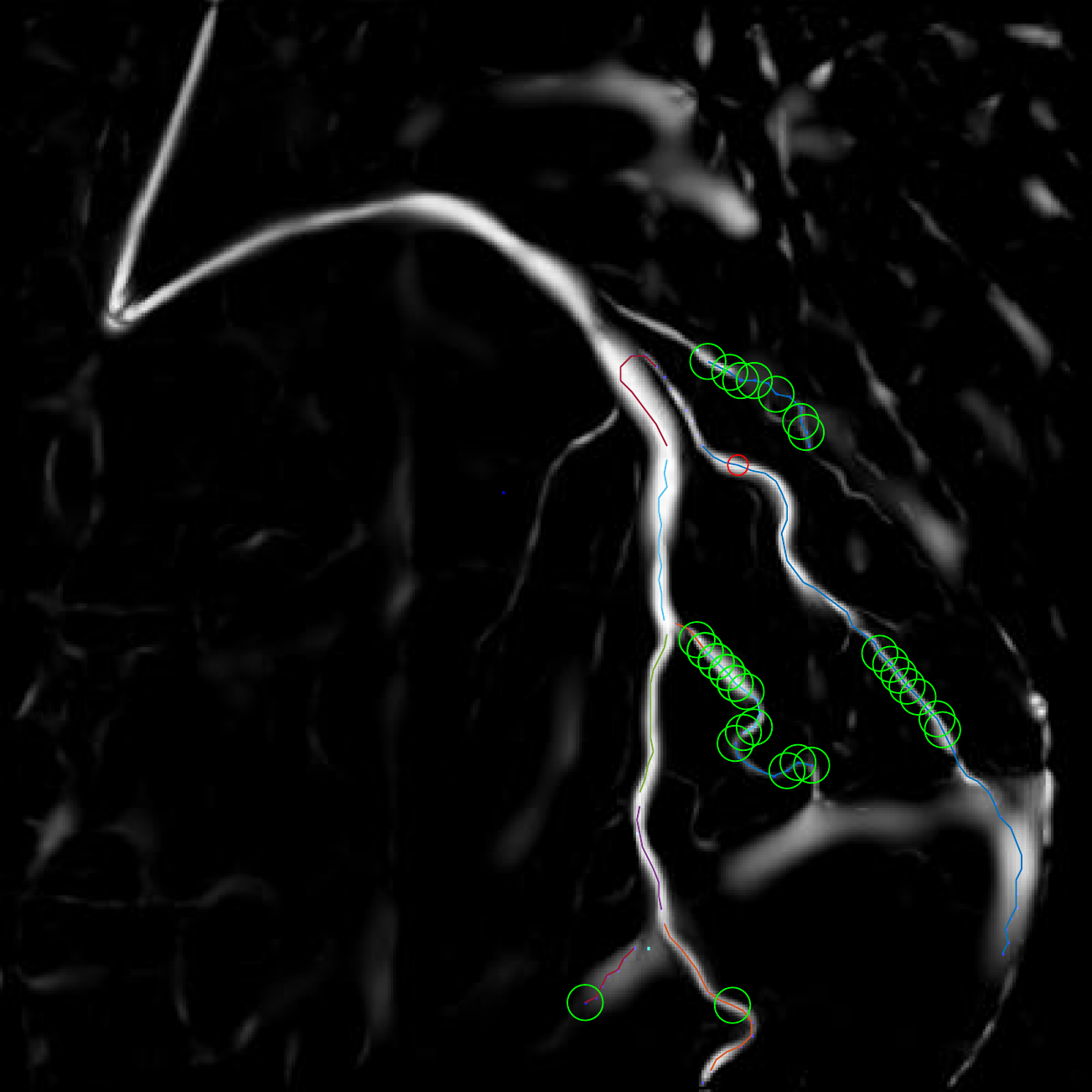} }
	\end{minipage}
	\begin{minipage}[t]{0.22\textwidth} 
		\subfigure[(a2)] { \label{fig6(a2)}     \includegraphics[width=1.2in]{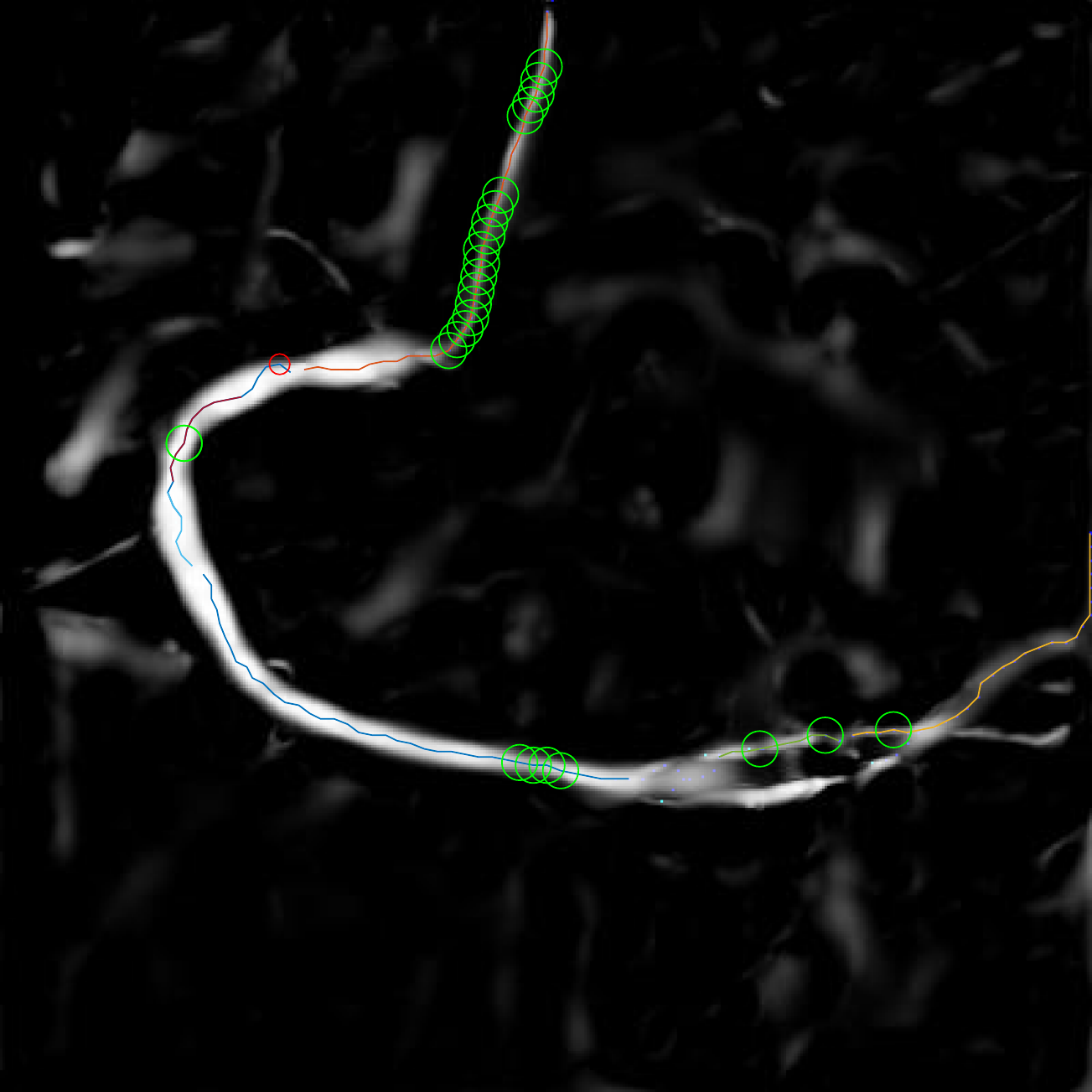} }   
	\end{minipage} 
	\begin{minipage}[t]{0.22\textwidth} 
		\subfigure[(a3)] { \label{fig7(a3)}     \includegraphics[width=1.2in]{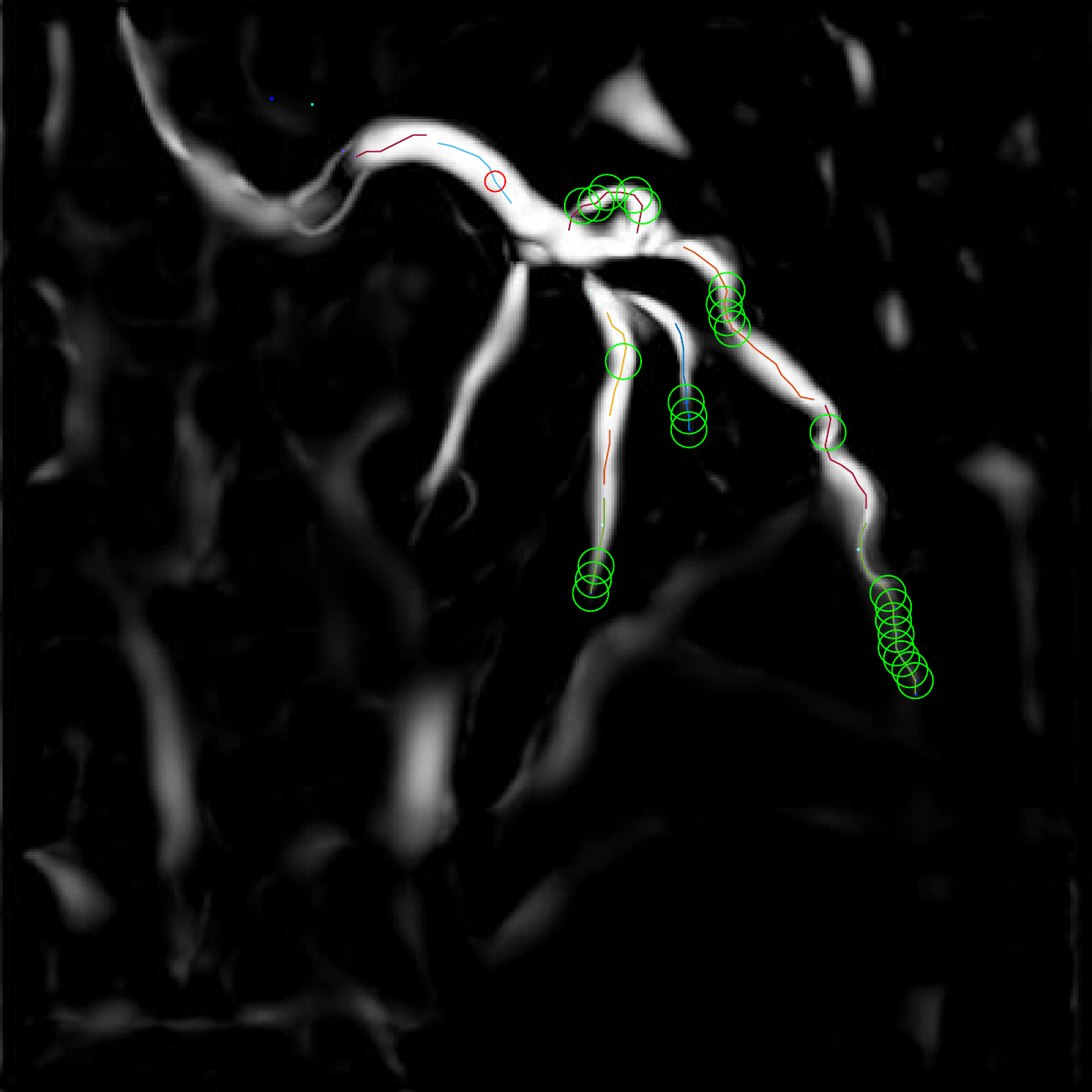} }   
	\end{minipage}
	\begin{minipage}[t]{0.22\textwidth} 
		\subfigure[(a4)] { \label{fig6(a4)}     \includegraphics[width=1.2in]{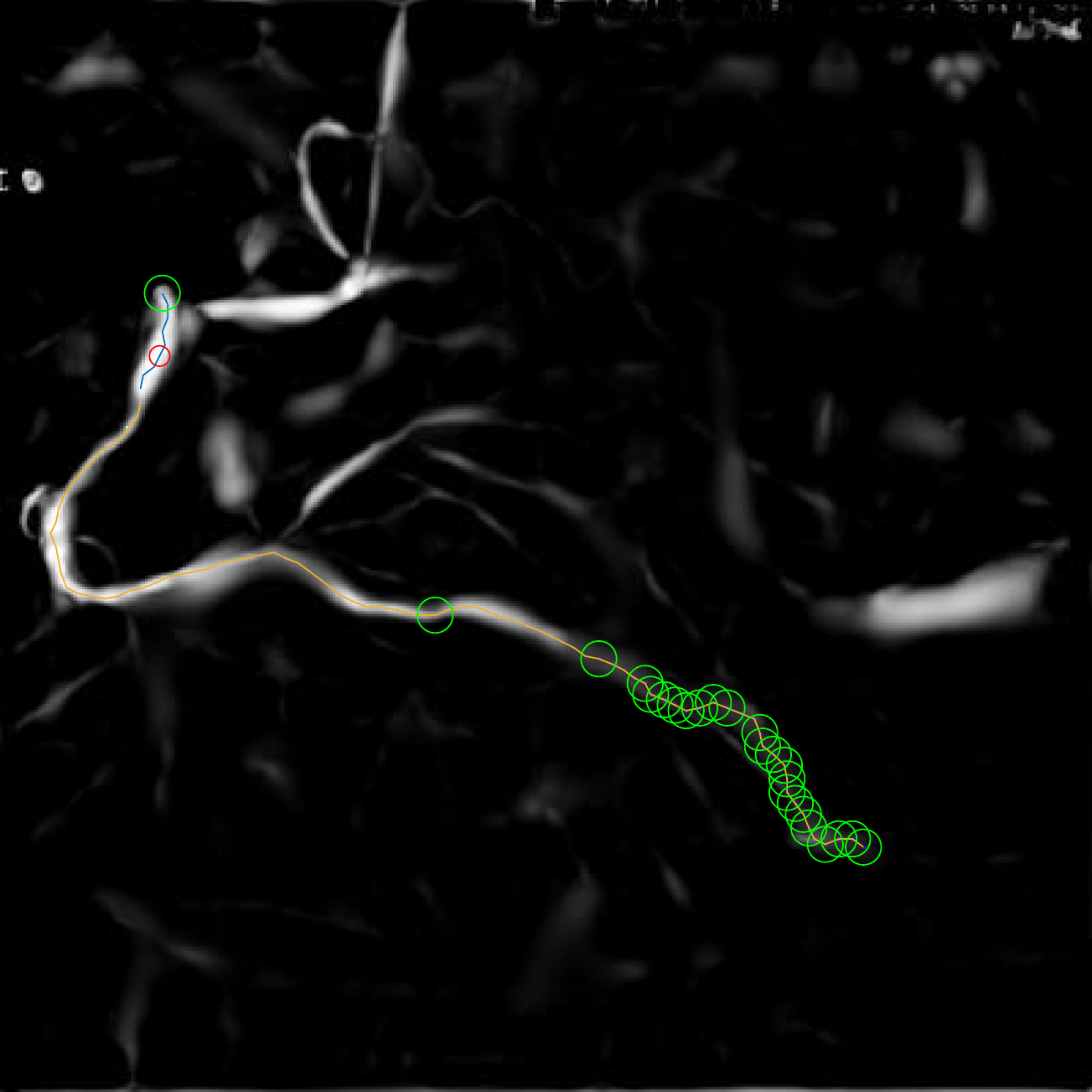} }   
	\end{minipage} 
	
	\begin{minipage}{0.22\textwidth}
		\subfigure[(b1)]{ \label{fig7(b1)}     \includegraphics[width=1.2in]{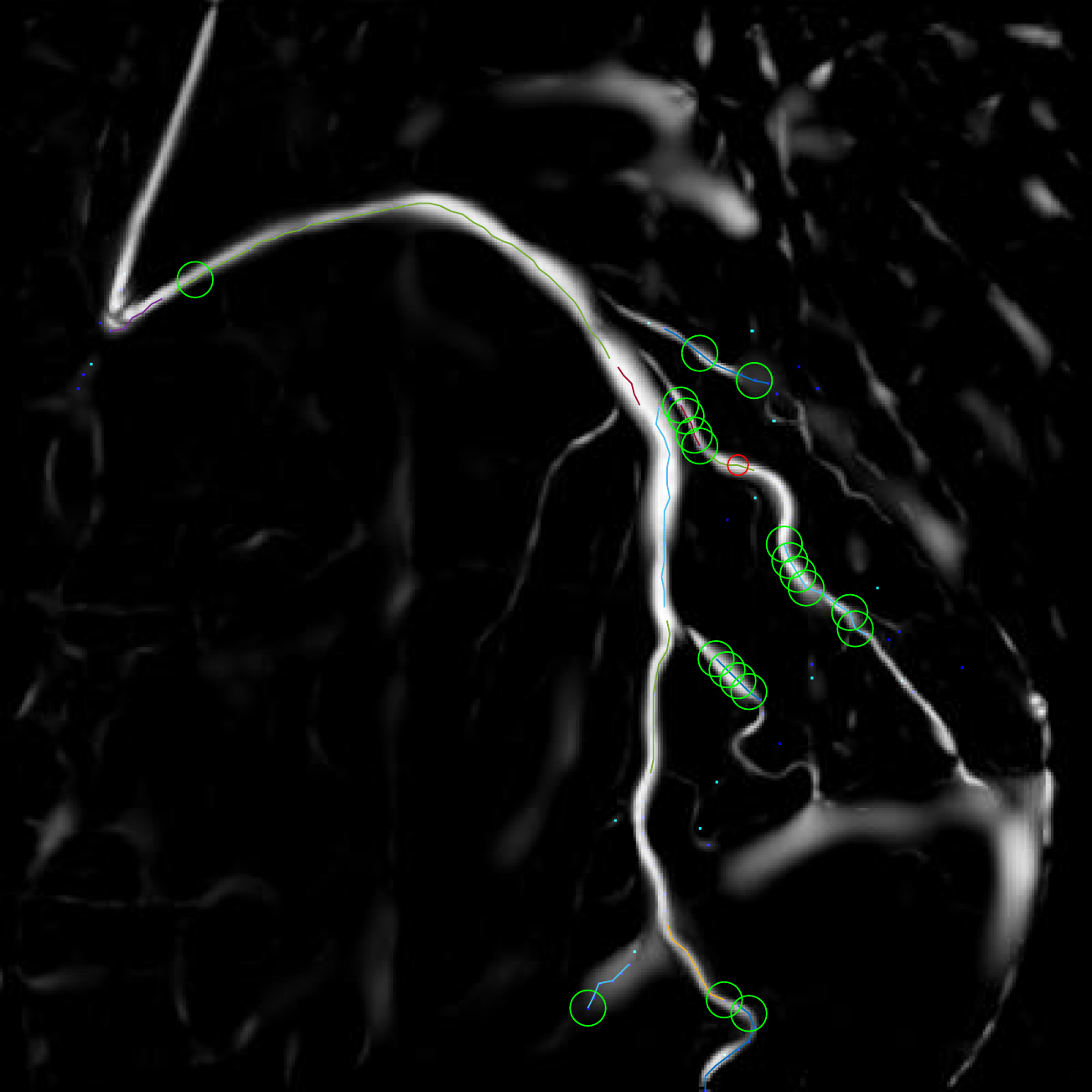} }
	\end{minipage}
	\begin{minipage}{0.22\textwidth} 
		\subfigure[(b2)] { \label{fig7(b2)}     \includegraphics[width=1.2in]{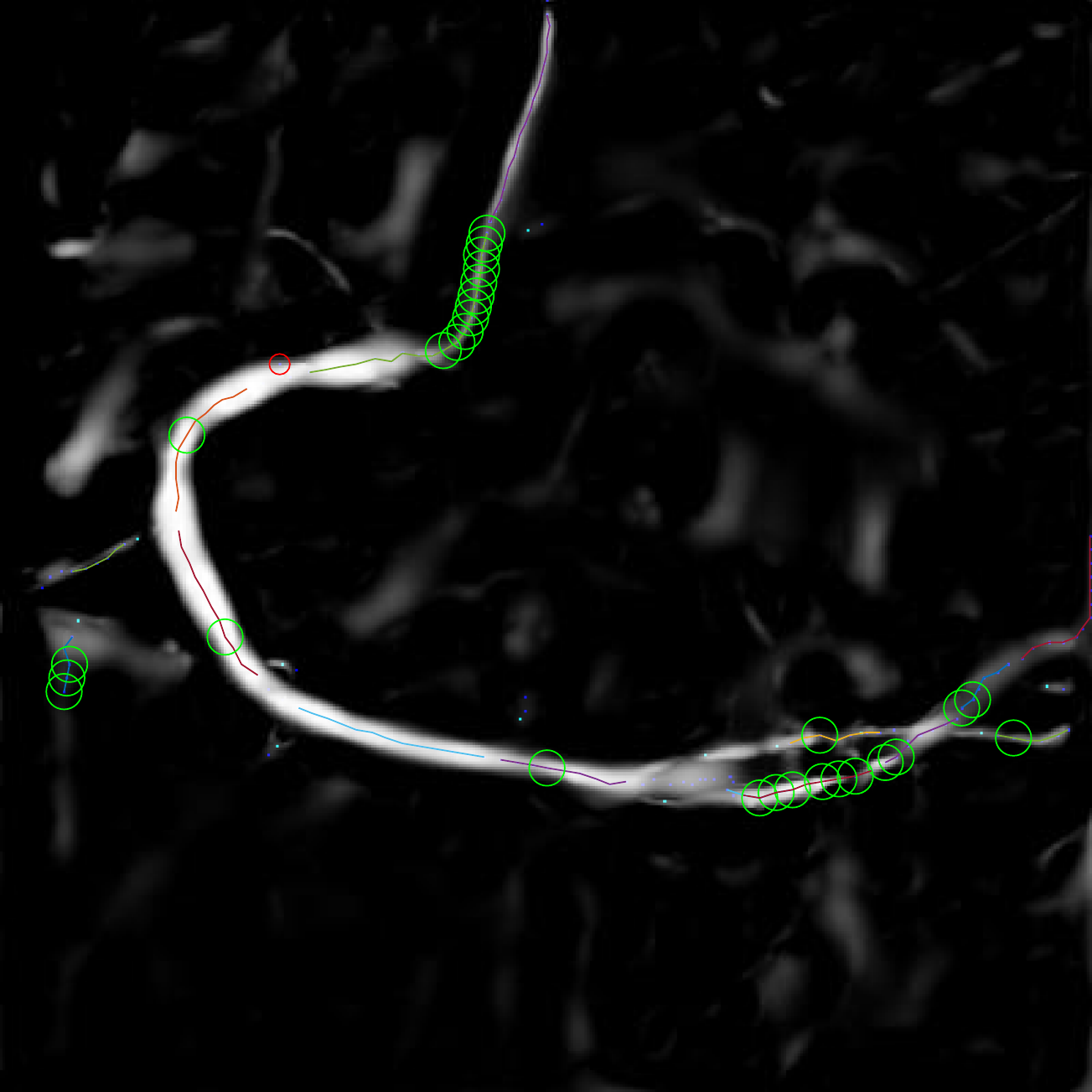} }   
	\end{minipage} 
	\begin{minipage}{0.22\textwidth}
		\subfigure[(b3)] { \label{fig7(b3)}     \includegraphics[width=1.2in]{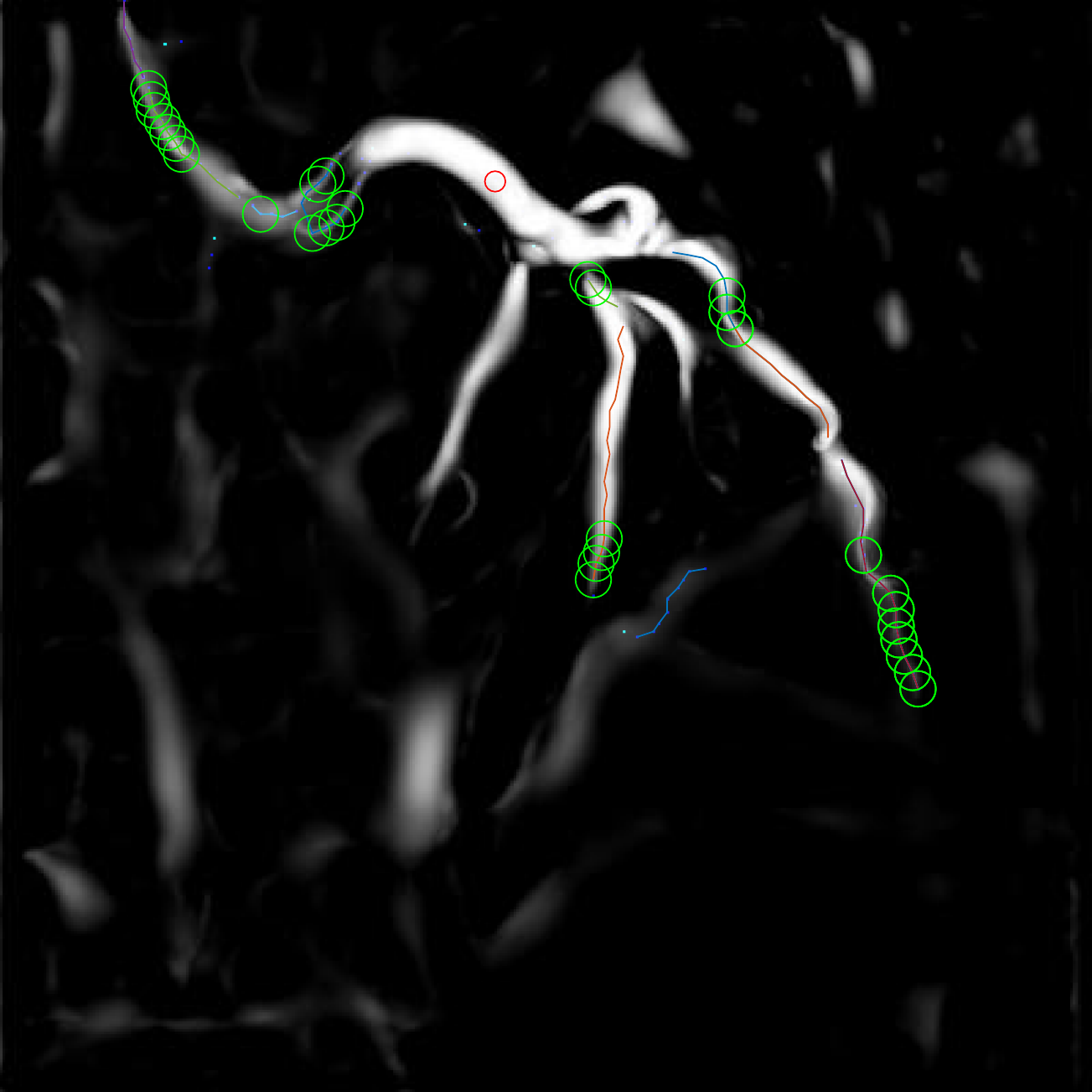} }   
	\end{minipage}
	\begin{minipage}{0.22\textwidth}
		\subfigure[(b4)] { \label{fig7(b4)}     \includegraphics[width=1.2in]{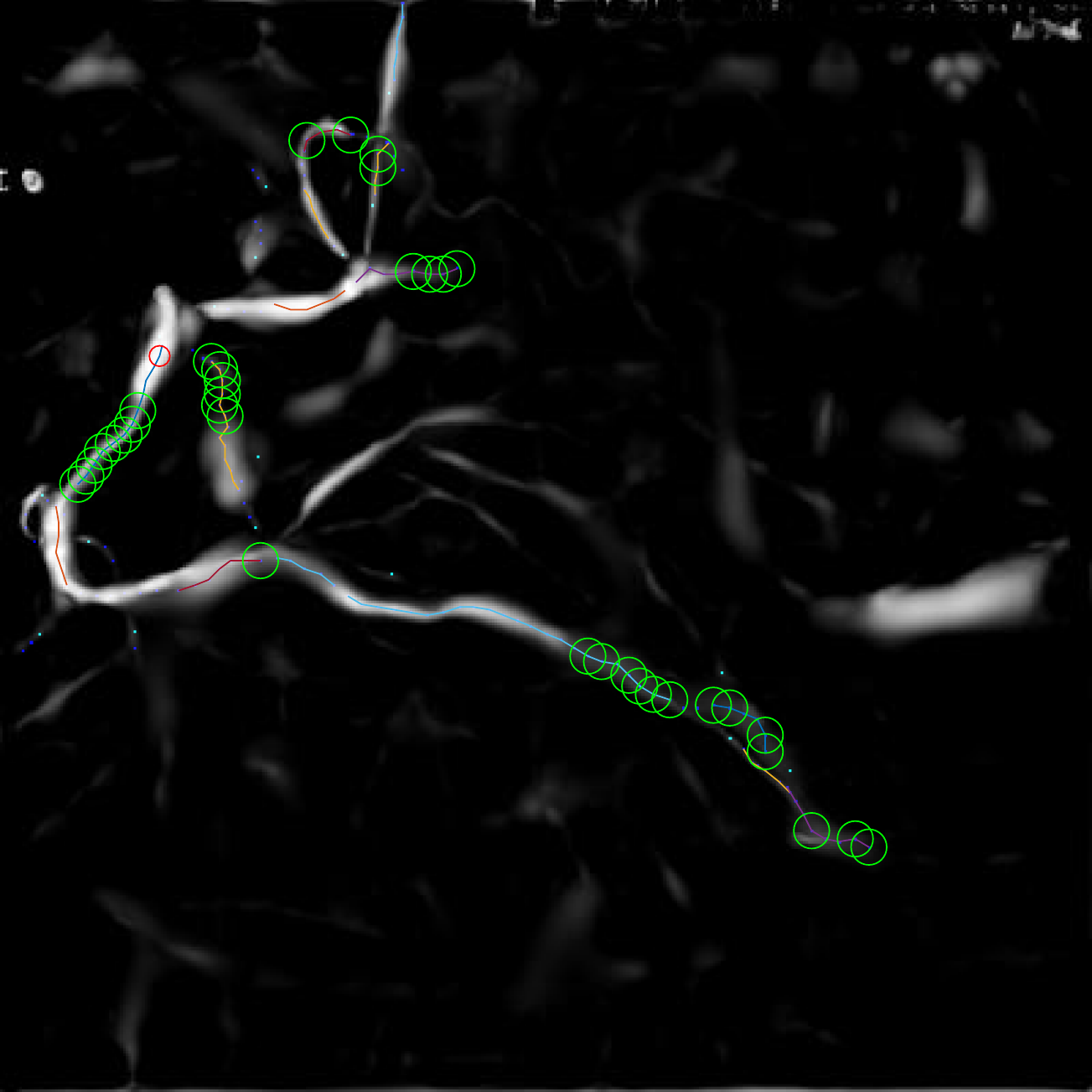} }   
	\end{minipage} 

	\begin{minipage}{0.22\textwidth}
		\subfigure[(c1)] { \label{fig7(c1)}     \includegraphics[width=1.2in]{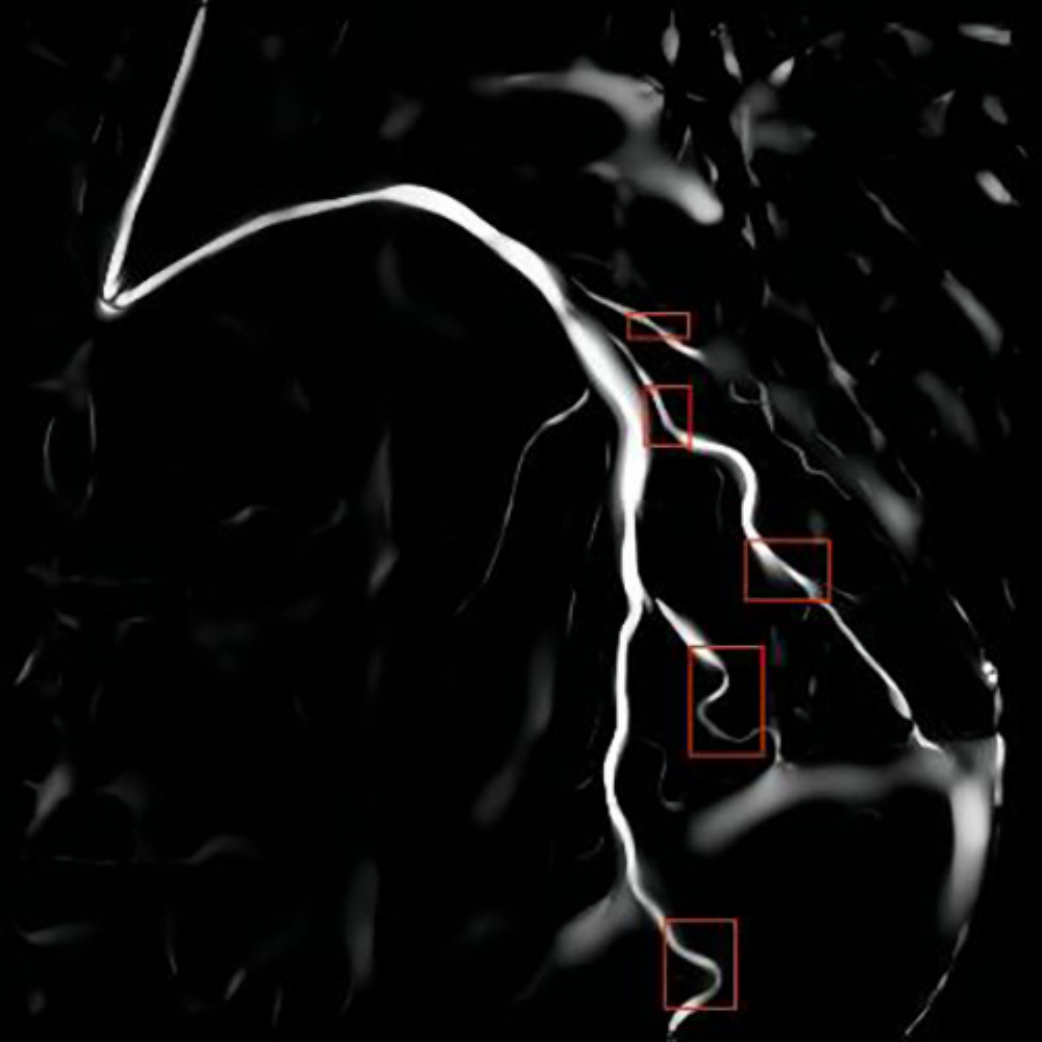} }
	\end{minipage}
	\begin{minipage}{0.22\textwidth}
		\subfigure[(c2)] { \label{fig7(c2)}     \includegraphics[width=1.2in]{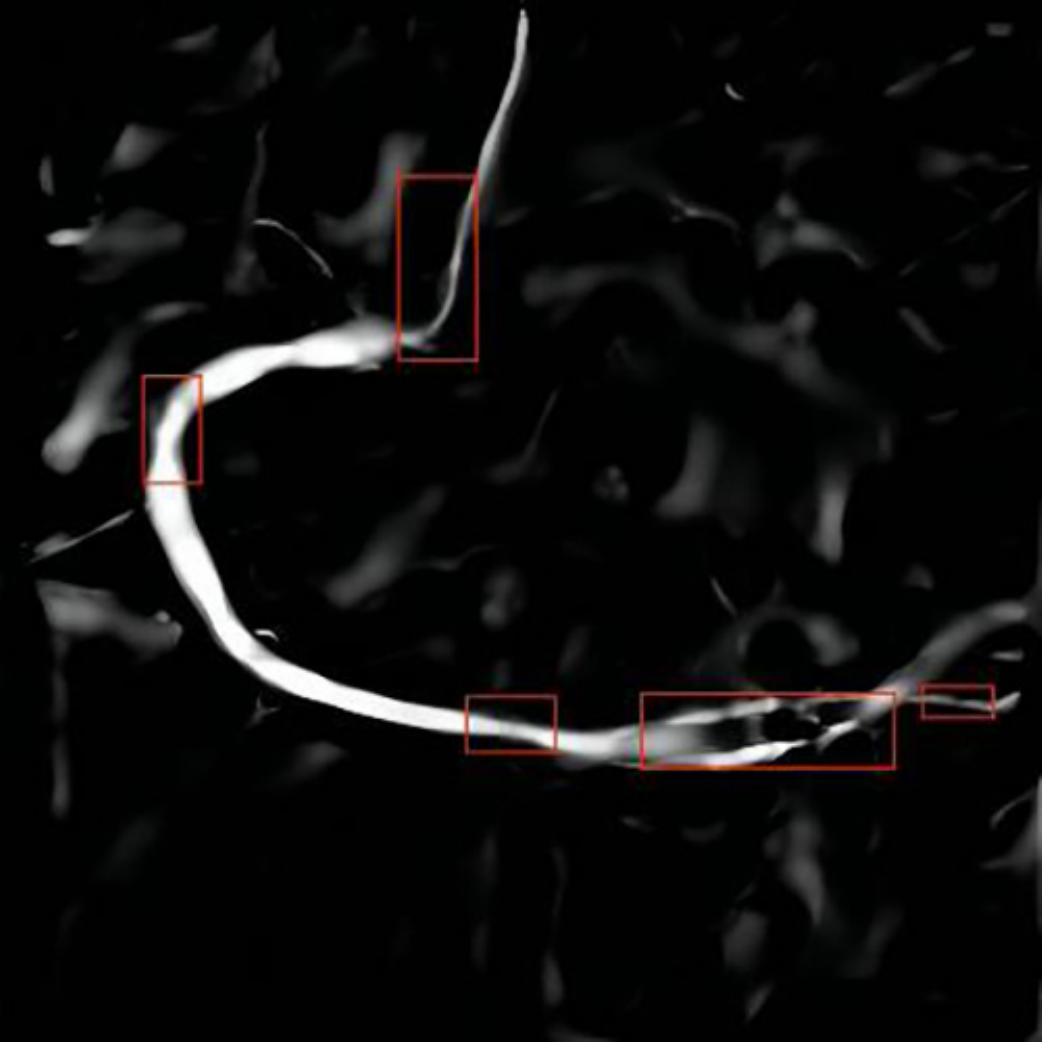} }   
	\end{minipage} 
	\begin{minipage}{0.22\textwidth}
		\subfigure[(c3)] { \label{fig7(c3)}     \includegraphics[width=1.2in]{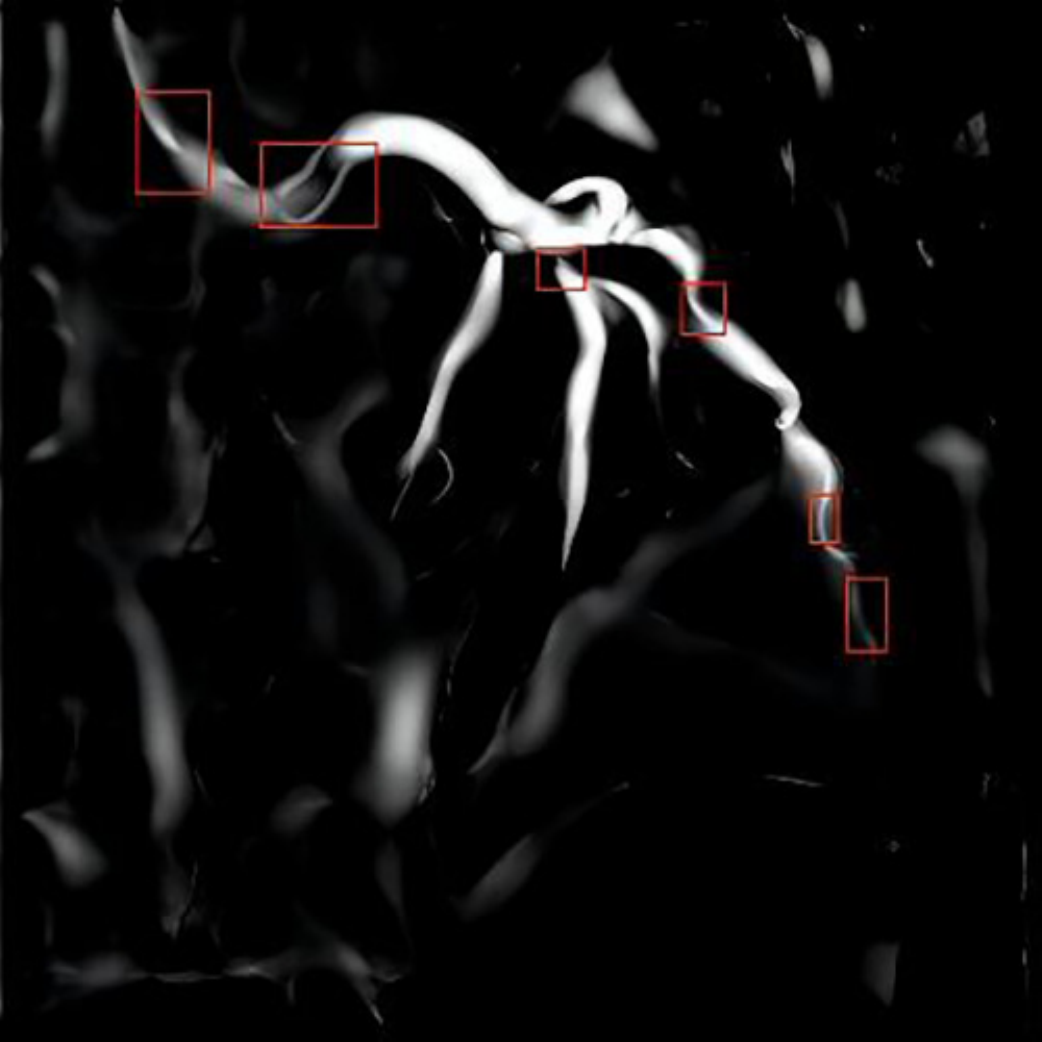} }   
	\end{minipage}
	\begin{minipage}{0.22\textwidth}
		\subfigure[(c4)] { \label{fig7(c4)}     \includegraphics[width=1.2in]{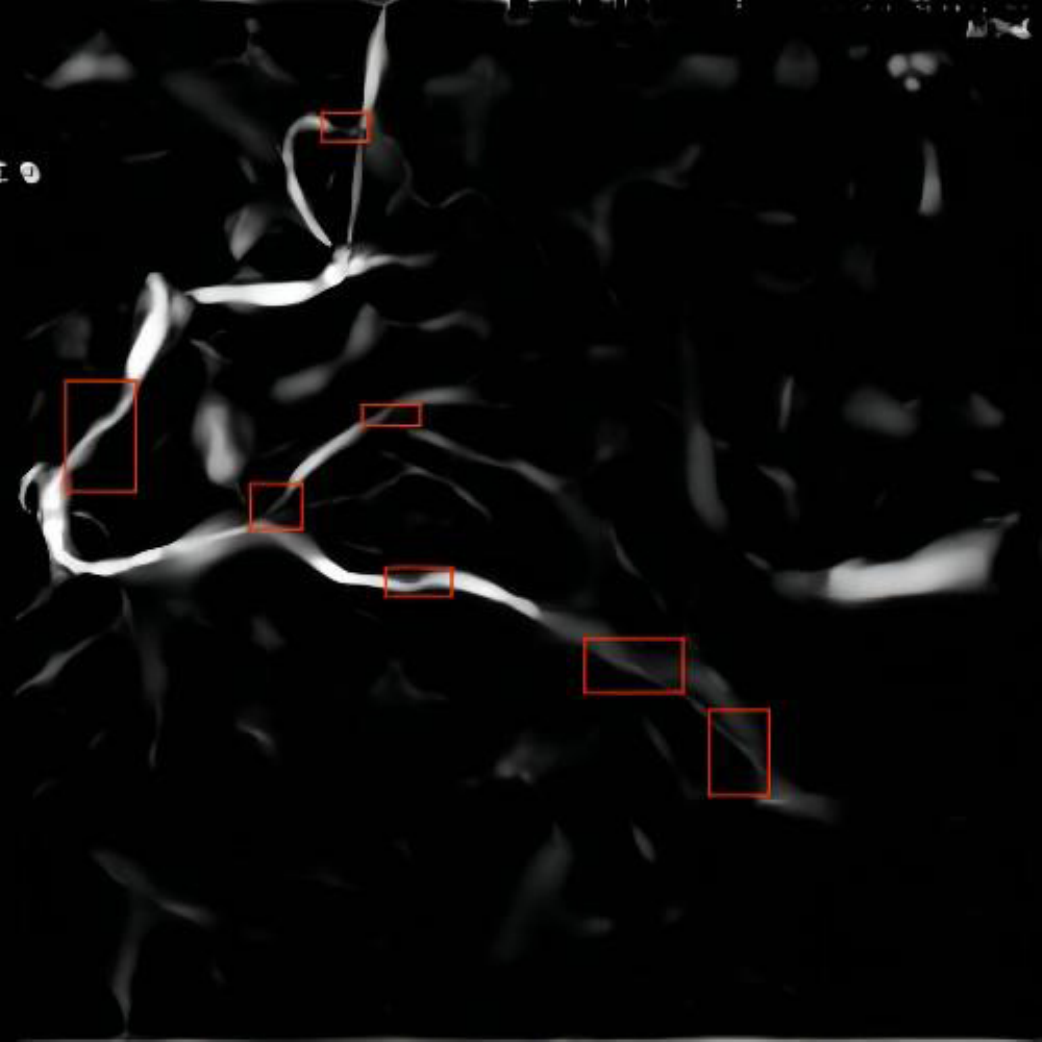} }   
	\end{minipage} 
	\centering
	\caption{Experimental results of automatic stenosis detection. (a) Xiao's method. (b) The proposed method. (c) Ground truth.}
	\label{Automatic results}
\end{figure*}

\subsection*{Interactive stenosis detection }
\label{sec3.3}
In practical applications, doctors may need to analyze a particular vessel segment of interest. Thus, we propose an interactive detection algorithm, which enables the user to detect and do quantitative analysis of stenoses on a particular segment by only interactively specifying a start point $P_{start}$ and an endpoint $P_{end}$. Precisely, we first extract the vessel segment between $P_{start}$ and  $P_{end}$ based on a proposed energy function. Then, we identify stenoses using our diameter measurement algorithm and Eq.\eqref{degree}\eqref{judge} mentioned above.

In extracting the segment, the complicated topology of the vessels is the most challenging  problem, especially the bifurcation structure, which can easily make the tracking point deviate from the correct route, i.e., it tracks to the other branch. To solve this problem, rather than using IAGVT directly, we introduce a potential energy function: 
\begin{equation}
	\begin{aligned}
		E\left(P\right)=I\left(P\right)+\frac{\lambda}{\sqrt{d\left(P,P_{end}\right)}},
	\end{aligned}
	\label{energy}
\end{equation}
where $\lambda$ is a coefficient. At every tracking step, we select the point with the largest value of the potential energy function on the search arc as the next tracking point so that we can get the correct tracking route. The first item makes the next tracking point not deviate from the vessel. The second item can make the tracking point avoid deviating at the bifurcation and gradually approaching  $P_{end}$. 

Using the above method, we significantly improved the effect of interactive detection. The specifical algorithm of interactive stenosis detection is as follows
\begin{enumerate}[(1)]
	\item Specify  $P_{start}$ and  $P_{end}$ interactively.
	\item Obtain the adjacent ridge points.  The start point $P'_{start}$ and the end point $P'_{end}$ can be formulated as:
	\begin{equation}
		\begin{cases}
			P'_{start}=\underset{P\in \mathcal{R}}{arg}\min d\left(P,P_{start}\right)\\
			P'_{end}=\underset{P\in \mathcal{R}}{arg}\min d\left(P,P_{end}\right)
		\end{cases},
	\end{equation}
	where $\mathcal{R}$ represents the set of the ridge points.
	\item Track. Take $P'_{start}$ as the seed point. The point $P_{j}$ with the largest value of the potential energy function Eq.\eqref{energy} on the search arc is selected as the next tracking point. Then, we determine whether the distance between the tracking point and $d(P_{j},P'_{end})$ is less than a given threshold $\tau_d$. If so, stop tracking. Otherwise, continue.
	\item Select the correct route. Considering there may be a ring structure, we perform step(3) in the forward and backward directions to get two routes and select the one with fewer tracking points as the correct route.
	\item Detect stenoses. We can identify stenoses at each tracking point using our algorithm of diameter measurement and Eq.\eqref{degree}\eqref{judge} mentioned above. 
	
\end{enumerate}

\begin{figure*}
	\centering
	\begin{minipage}{0.22\textwidth}
		\subfigure[(a)] { \label{fig8(a)}     \includegraphics[width=1.2in]{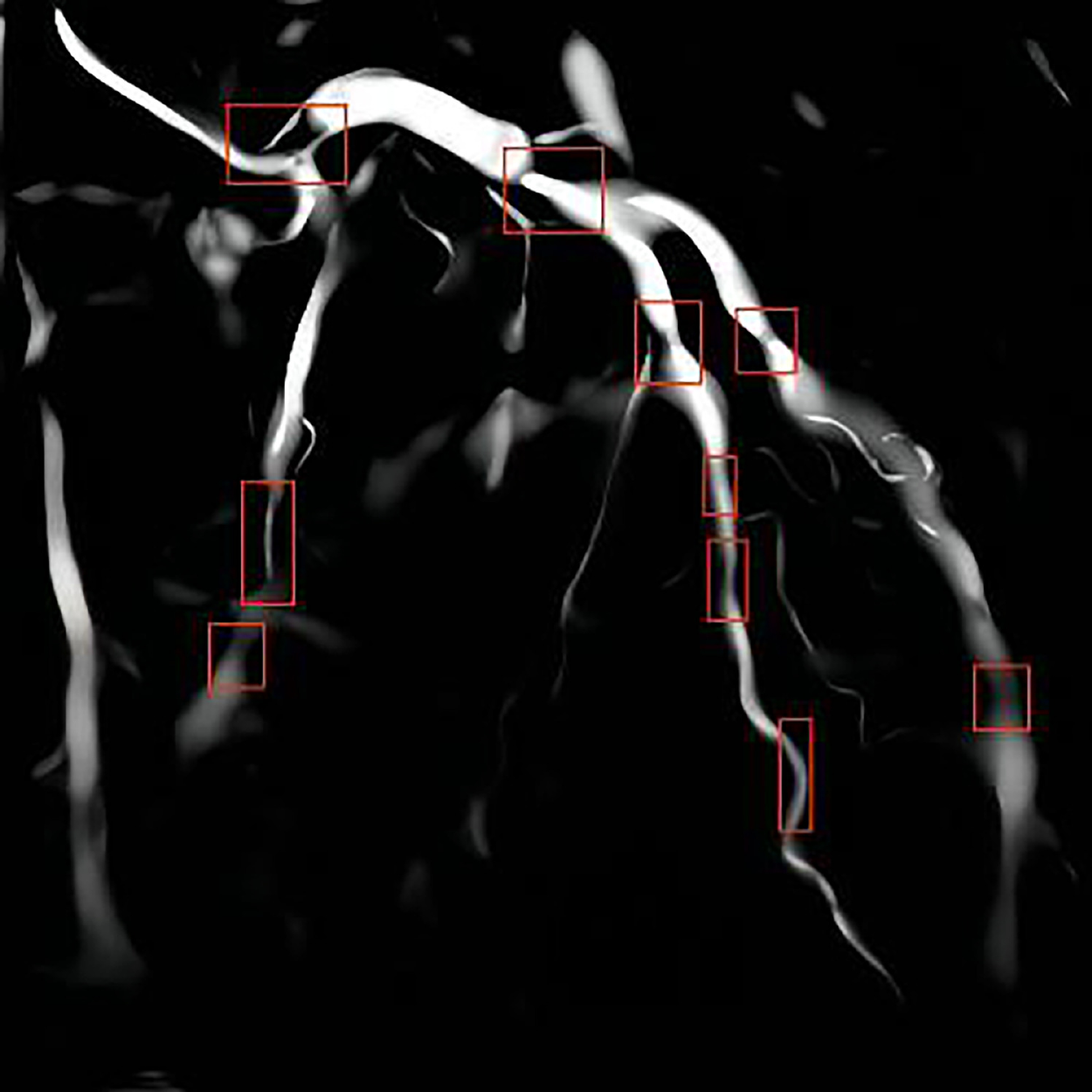} }
	\end{minipage}
	\begin{minipage}{0.22\textwidth} 
		\subfigure[(b)] { \label{fig8(b)}     \includegraphics[width=1.2in]{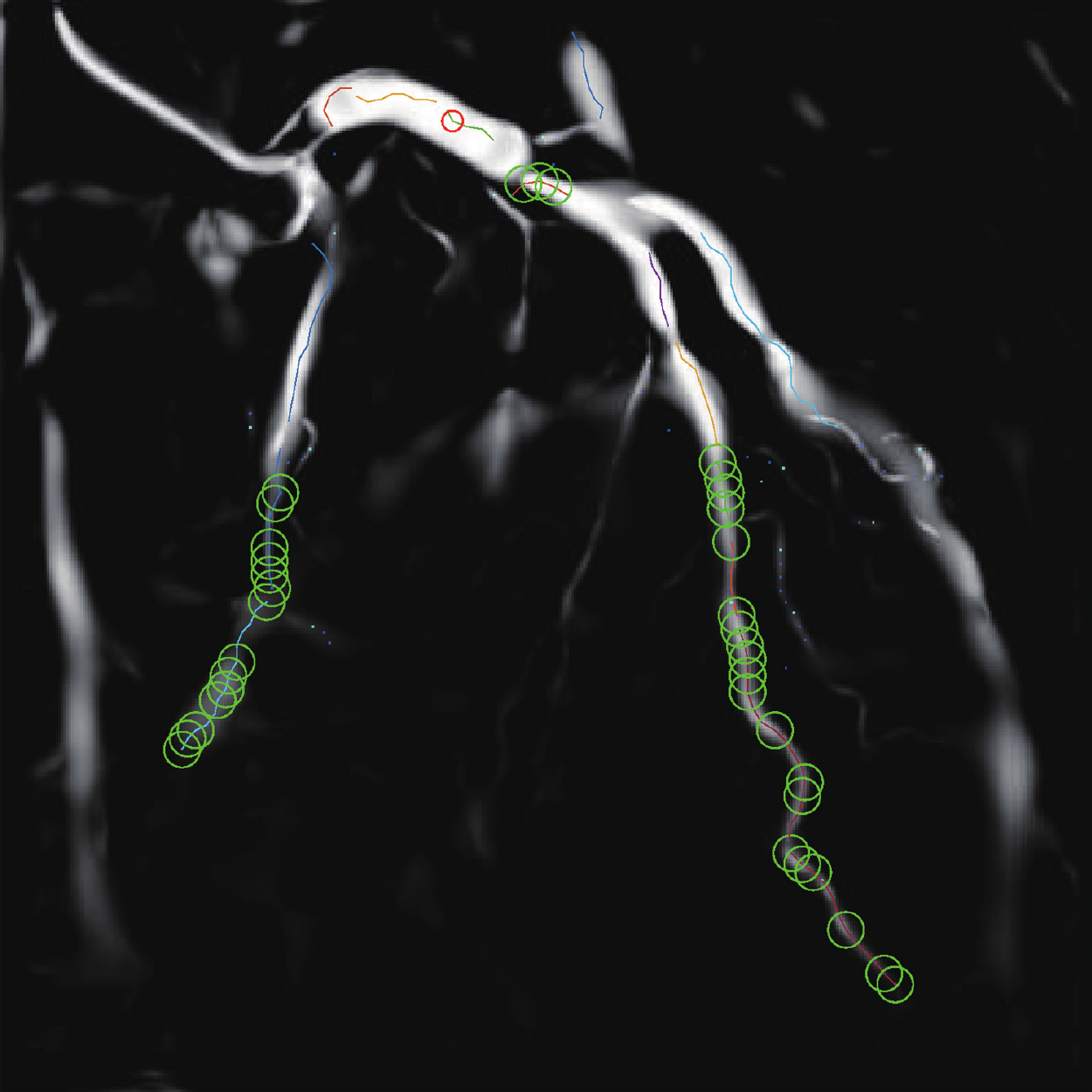} }
	\end{minipage} 
	\begin{minipage}{0.22\textwidth}
		\subfigure[(c1)] { \label{fig8(c1)}     \includegraphics[width=1.2in]{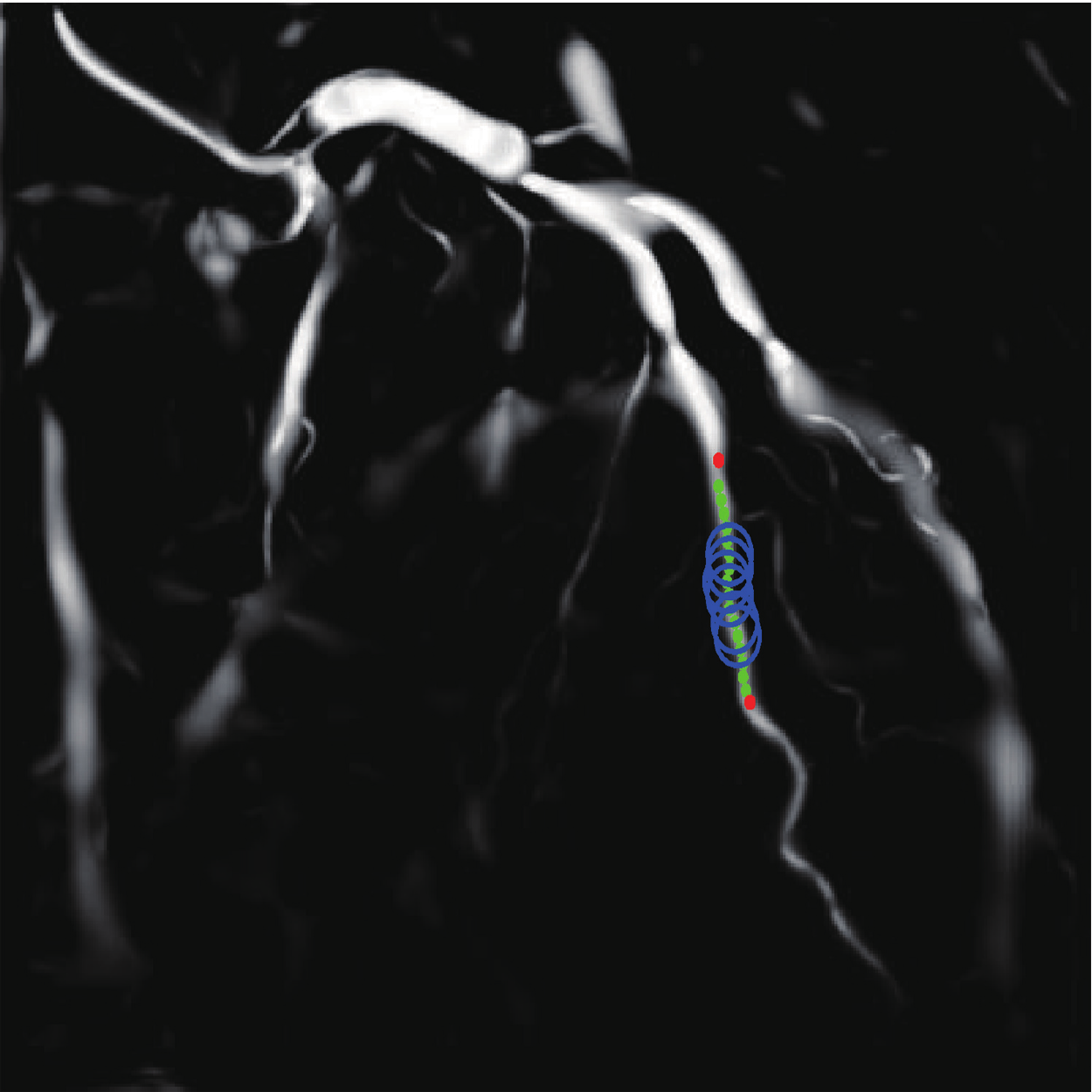} }
	\end{minipage}
	\begin{minipage}{0.22\textwidth} 
		\subfigure[(c2)] { \label{fig8(c2)}     \includegraphics[width=1.2in]{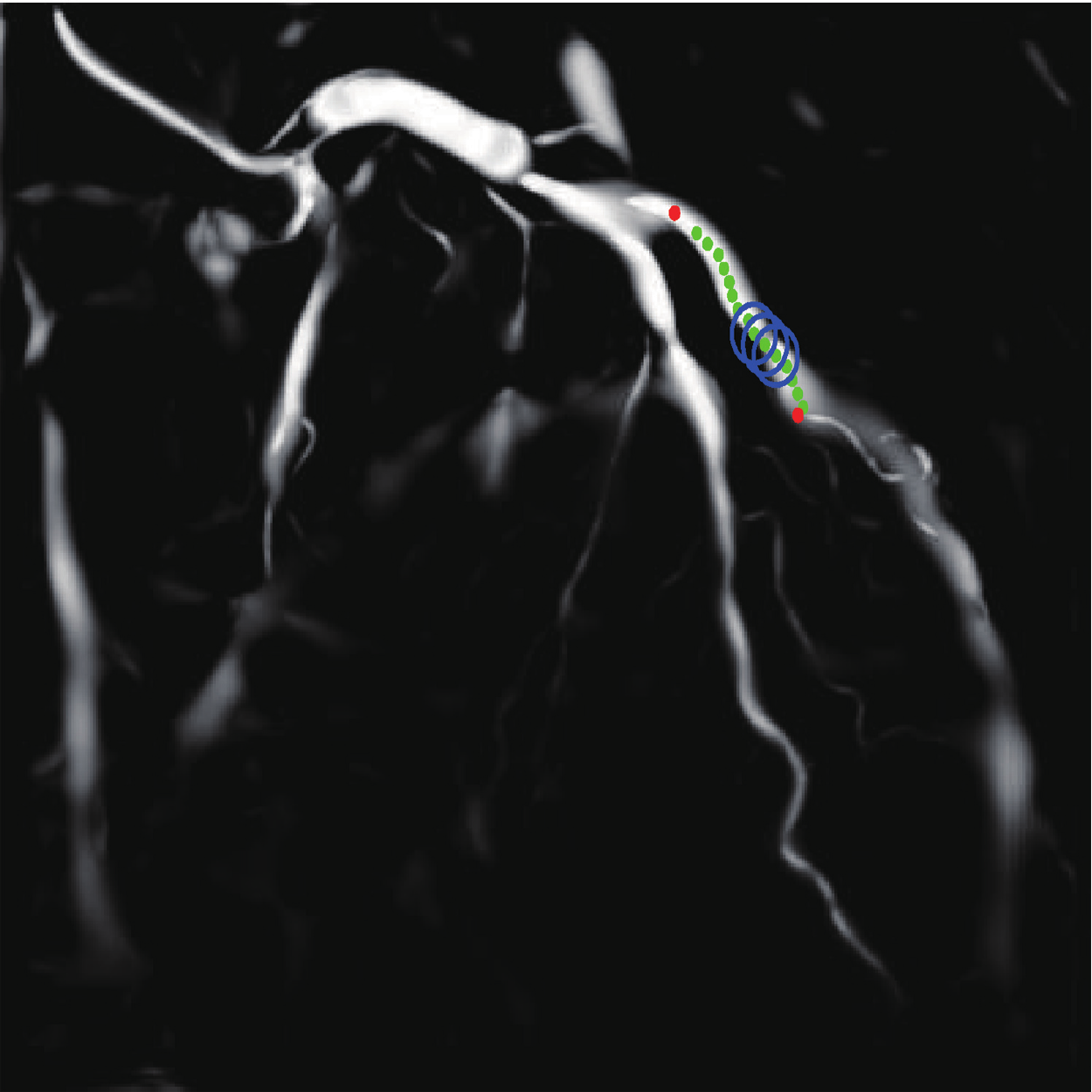} }
	\end{minipage} 
	
	\centering
	\caption{Results of more precise stenosis detection using the interactive method. (a) Ground truth. (b) Automatic method. (c) Interactive method.}
	\label{Interactive results}
\end{figure*}

\begin{figure*}[t]
	\centering
	\begin{minipage}[t]{0.3\textwidth}
		\subfigure[(a1)] { \label{fig9(a1)}     \includegraphics[width=1.5in]{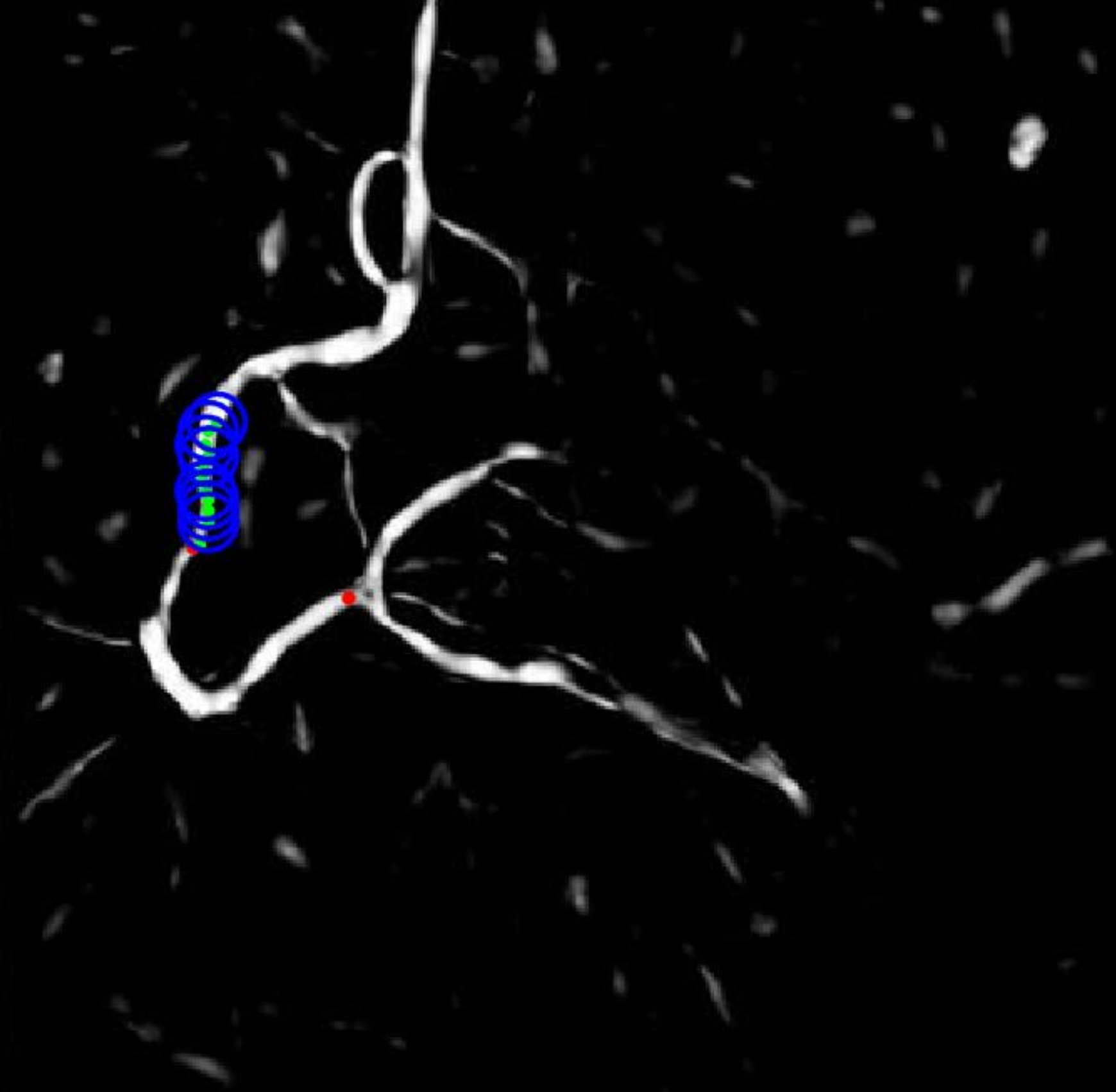} }
	\end{minipage}
	\begin{minipage}[t]{0.3\textwidth} 
		\subfigure[(a2)] { \label{fig9(a2)}     \includegraphics[width=1.5in]{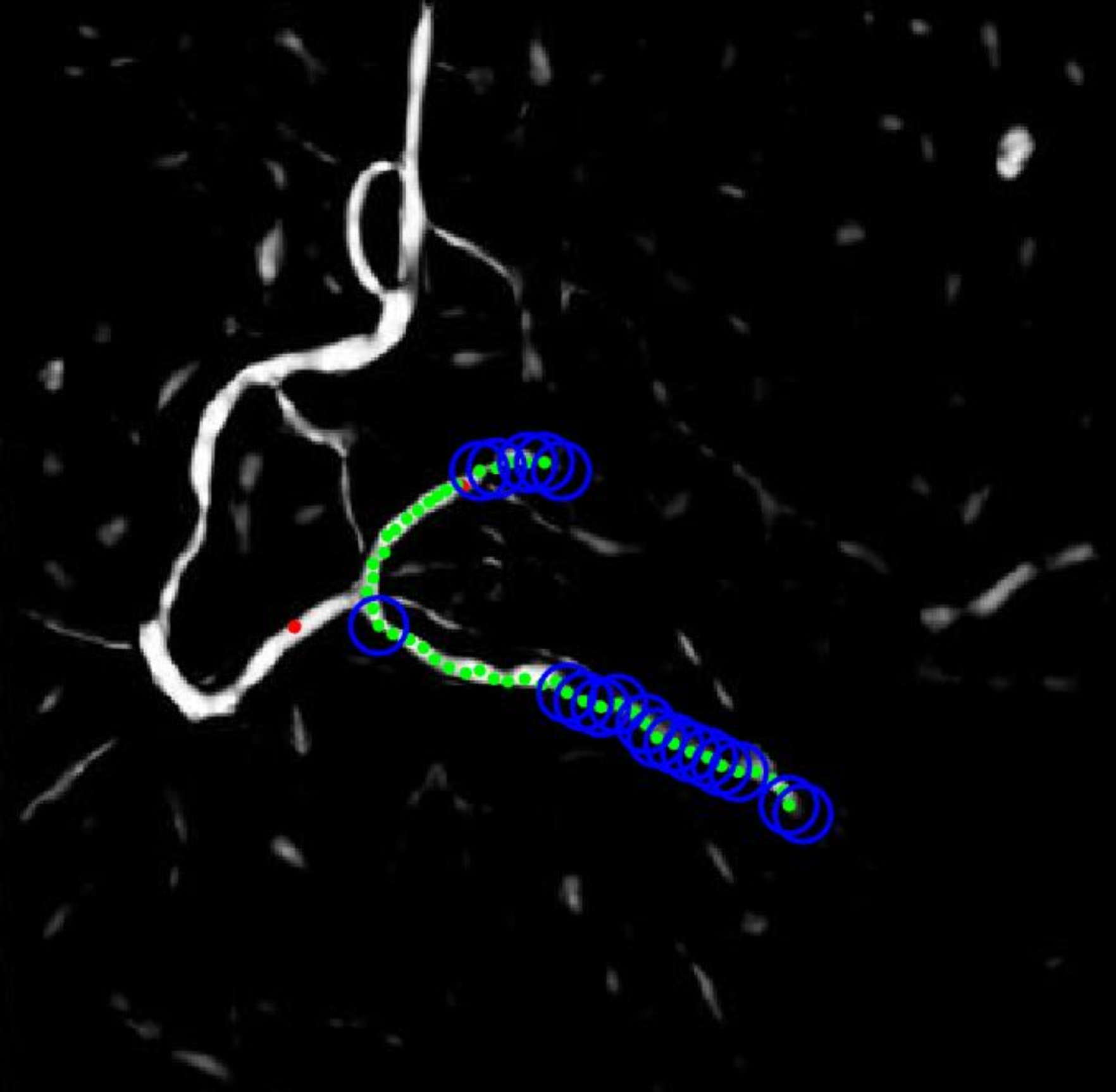} }   
	\end{minipage} 
	\begin{minipage}[t]{0.3\textwidth} 
		\subfigure[(a3)] { \label{fig9(a3)}     \includegraphics[width=1.5in]{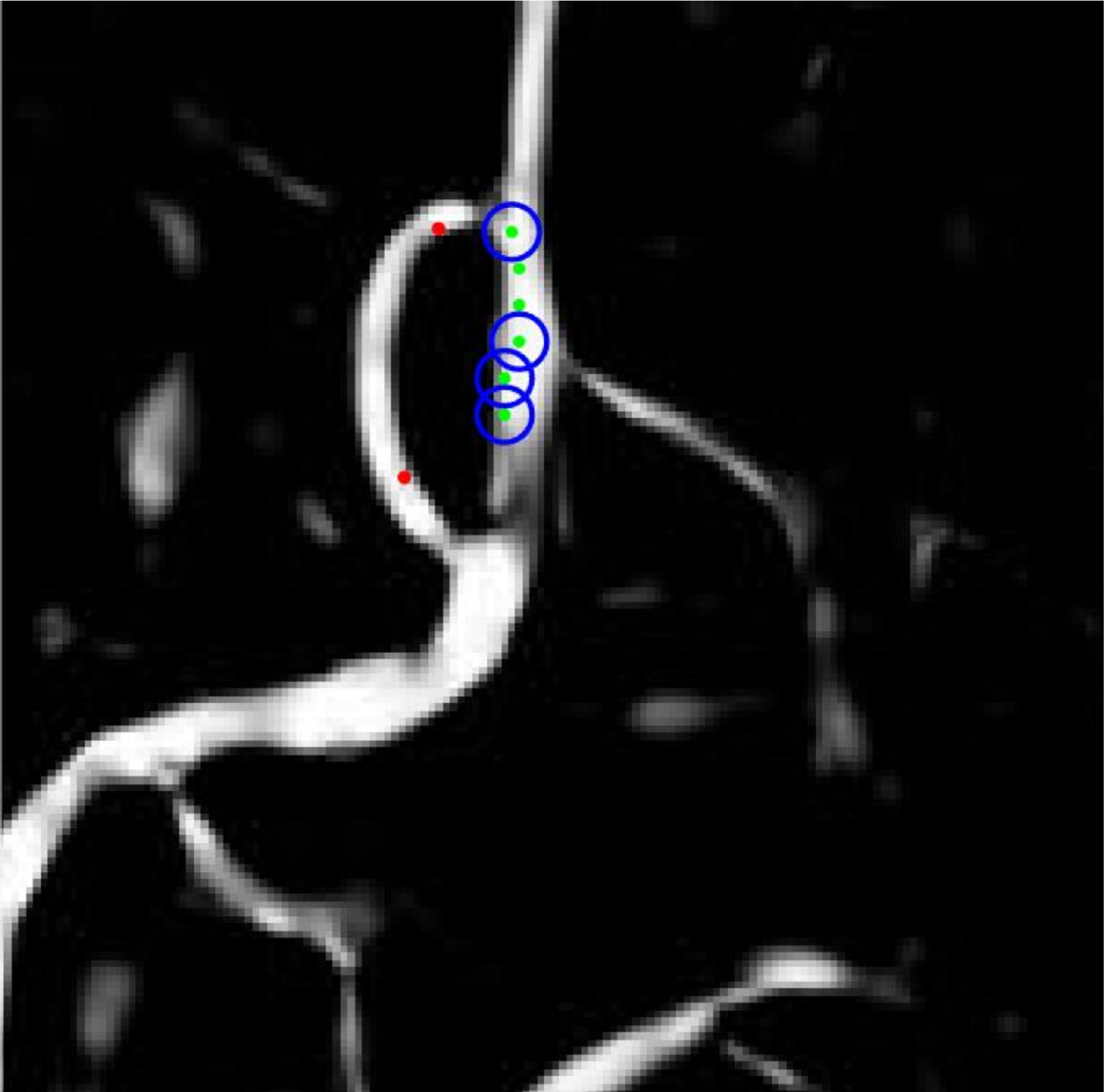} }   
	\end{minipage}
	
	\begin{minipage}{0.3\textwidth}
		\subfigure[(b1)] { \label{fig9(b1)}     \includegraphics[width=1.5in]{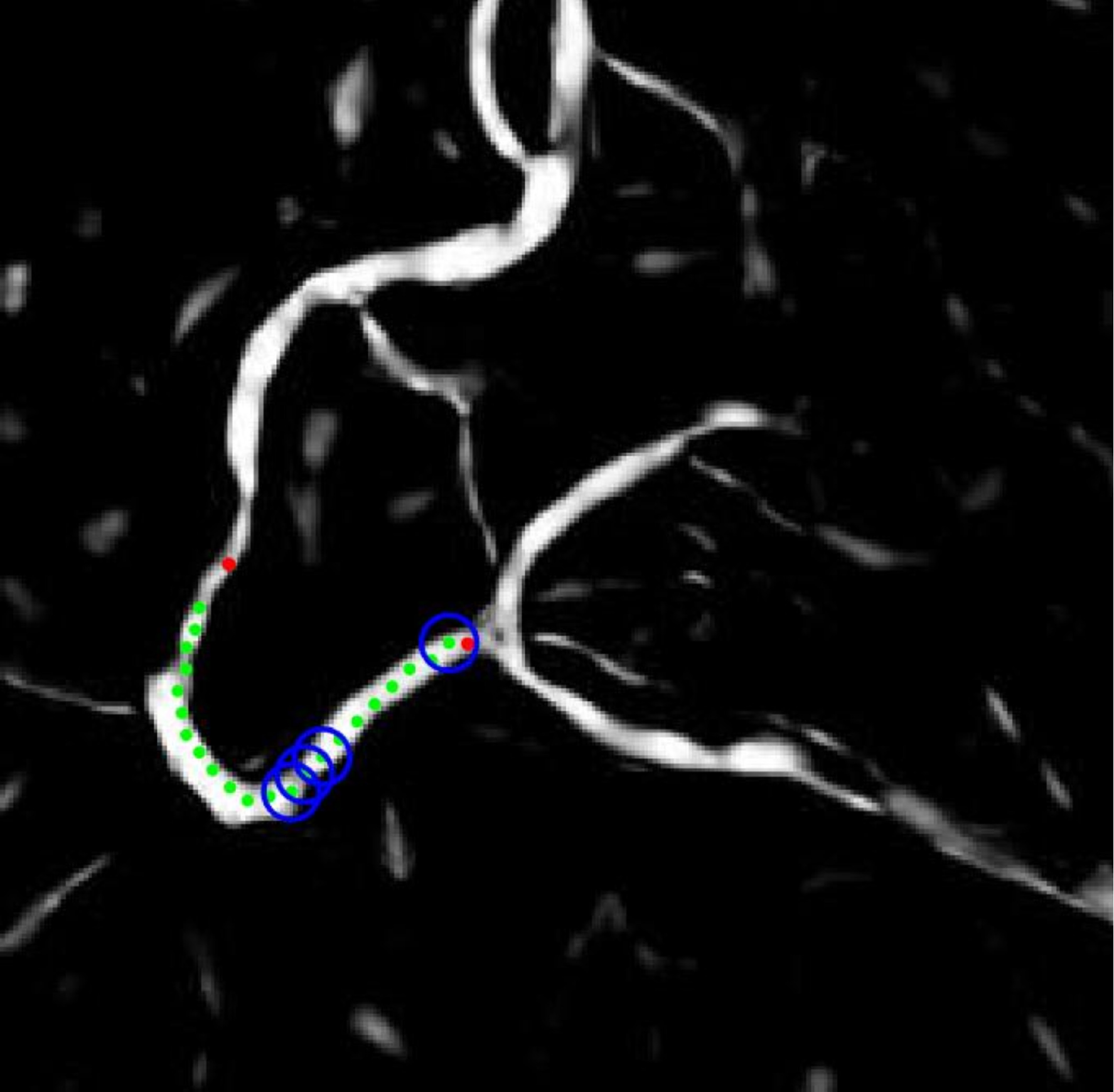} }
	\end{minipage}
	\begin{minipage}{0.3\textwidth} 
		\subfigure[(b2)] { \label{fig9(b2)}     \includegraphics[width=1.5in]{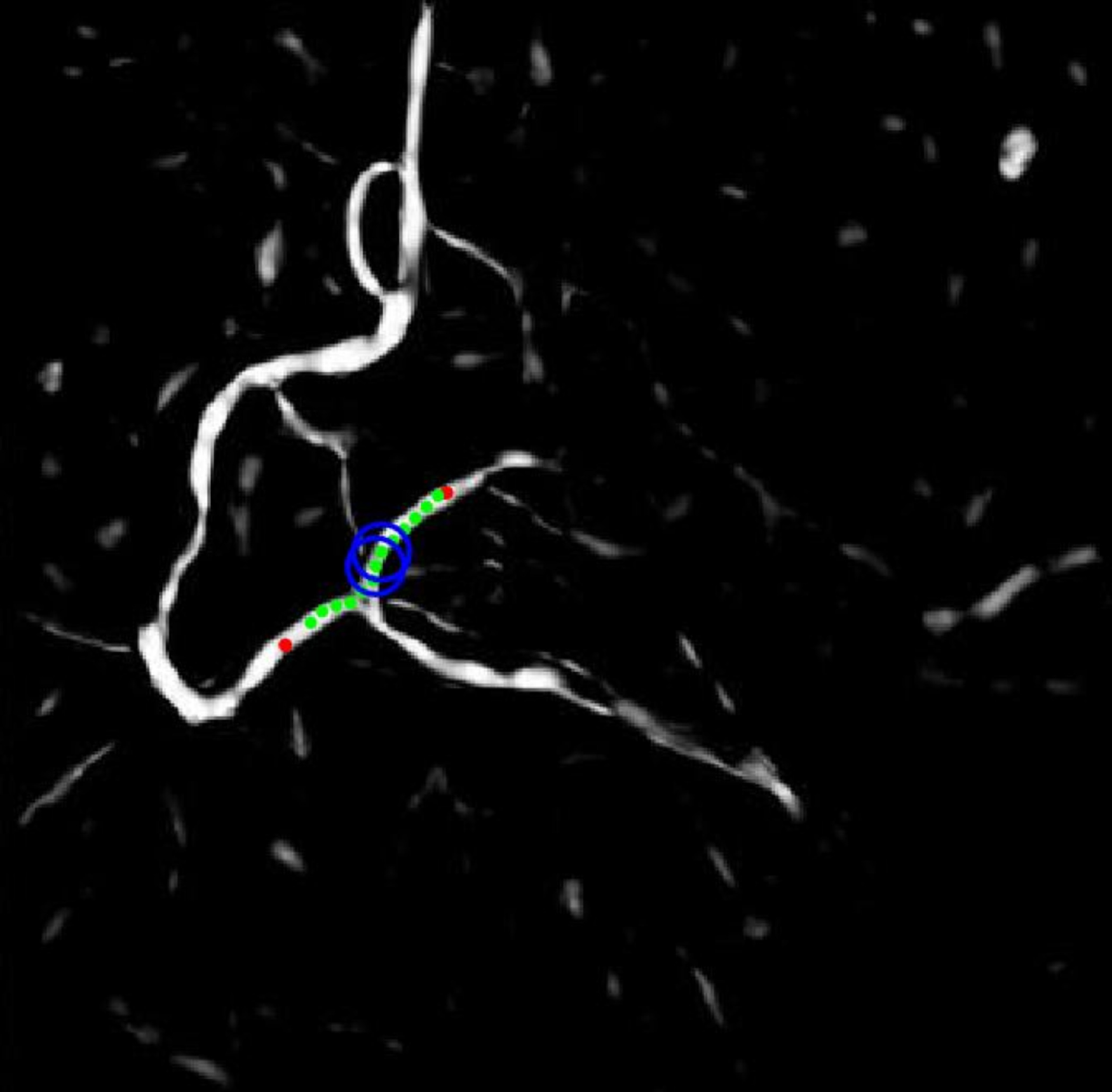} }   
	\end{minipage} 
	\begin{minipage}{0.3\textwidth} 
		\subfigure[(b3)] { \label{fig9(b3)}     \includegraphics[width=1.5in]{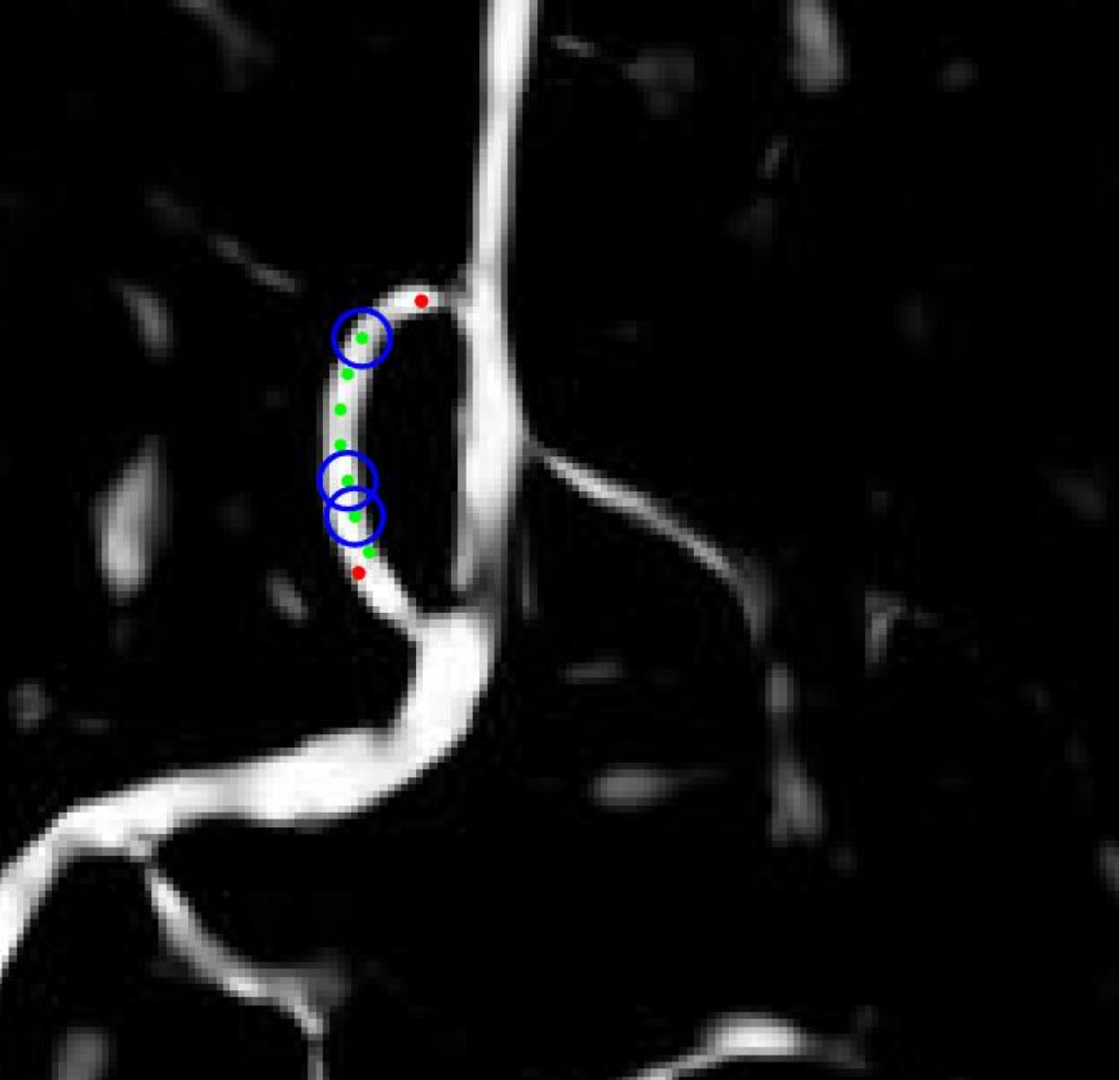} }   
	\end{minipage}
	
	\centering
	\caption{Comparison of the interactive stenosis detection before ((a)) and after ((b)) the introduction of the potential energy function}
	\label{Comparison interactive}
\end{figure*}

\begin{figure*}[t]
	\centering
	\subfigure[(a1)] { \label{fig10(a1)}     \includegraphics[width=1.15in]{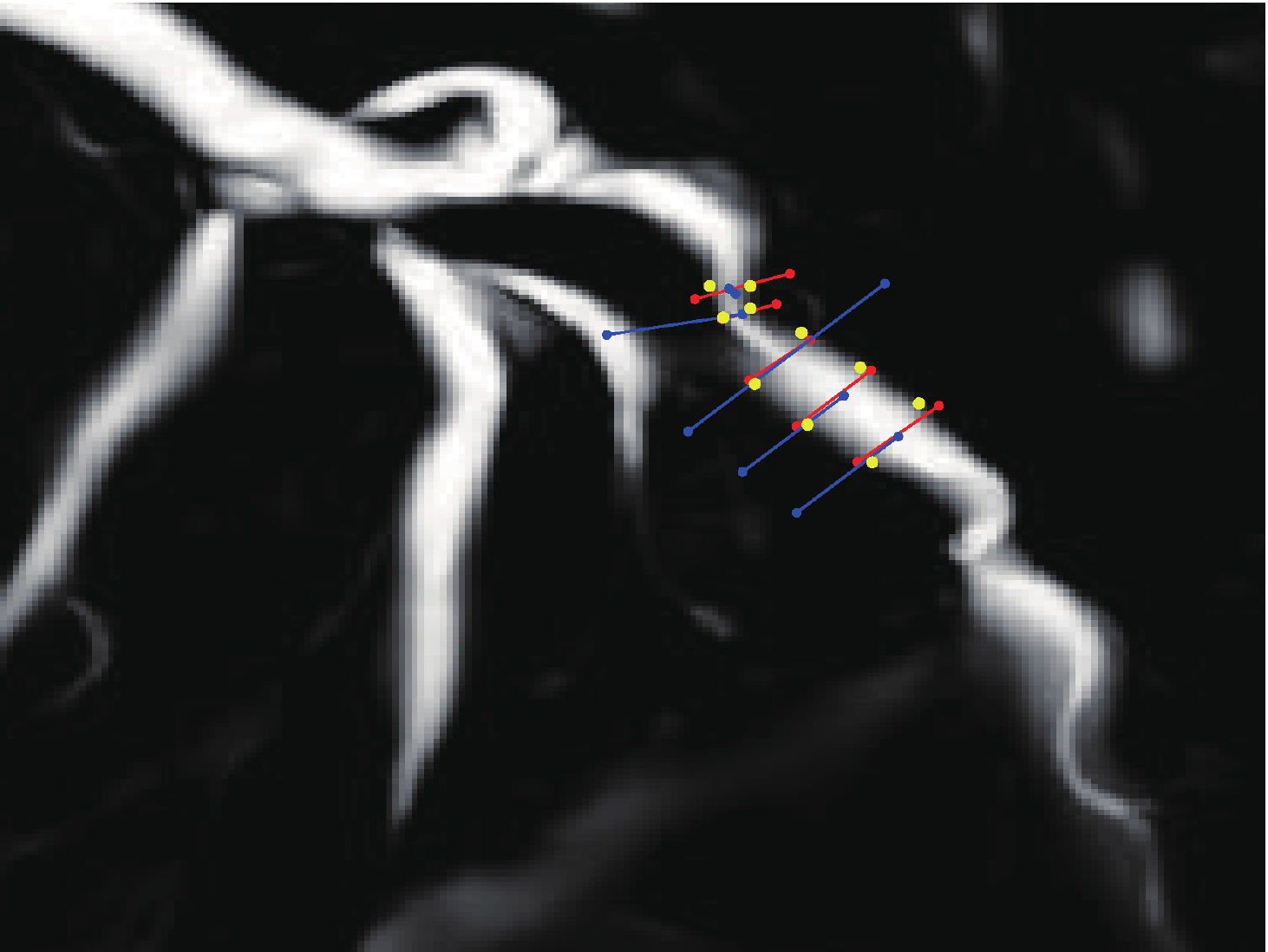} }
	\subfigure[(a2)] { \label{fig10(a2)}     \includegraphics[width=1.15in]{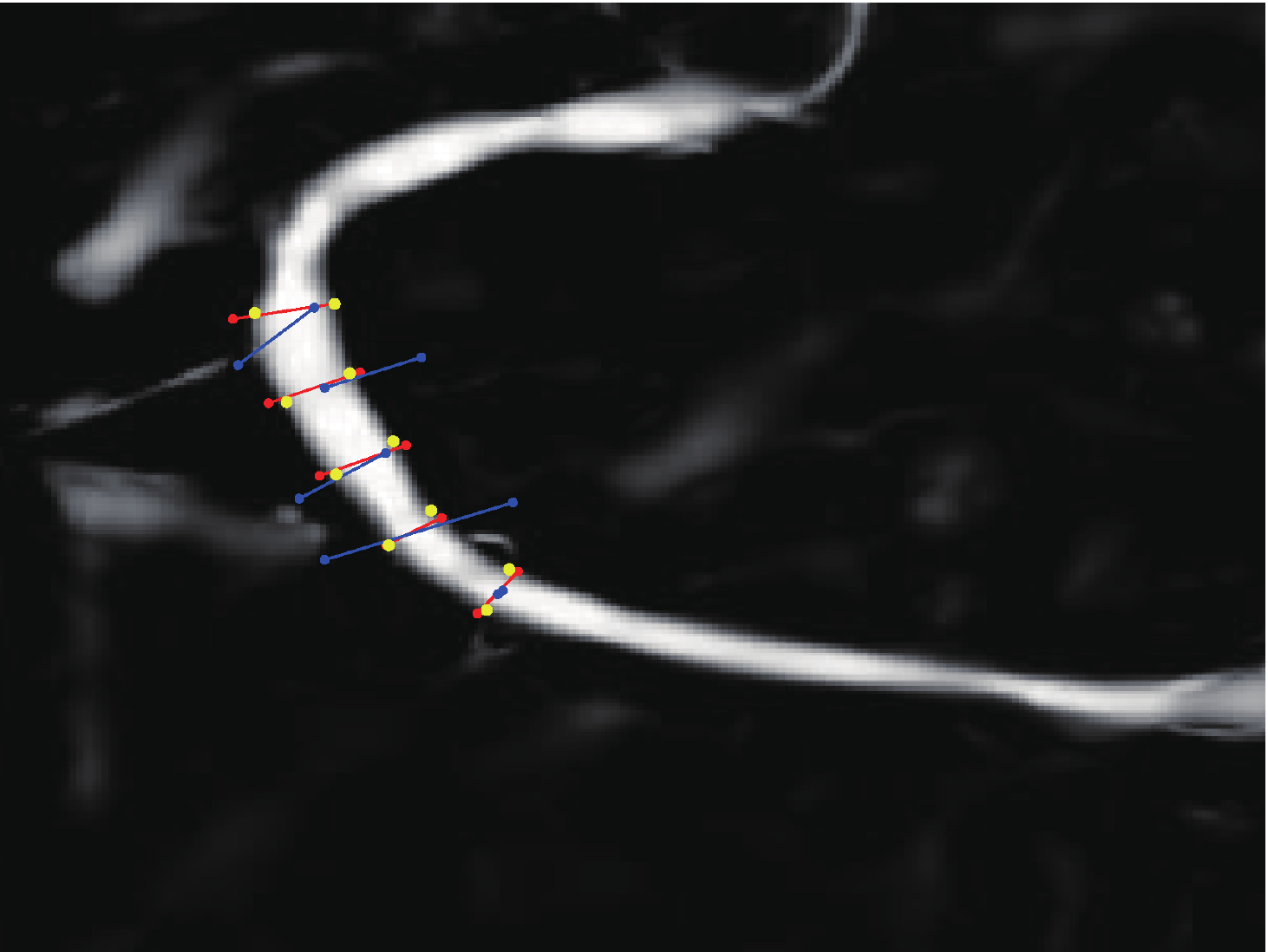} }   
	\subfigure[(a3)] { \label{fig10(a3)}     \includegraphics[width=1.15in]{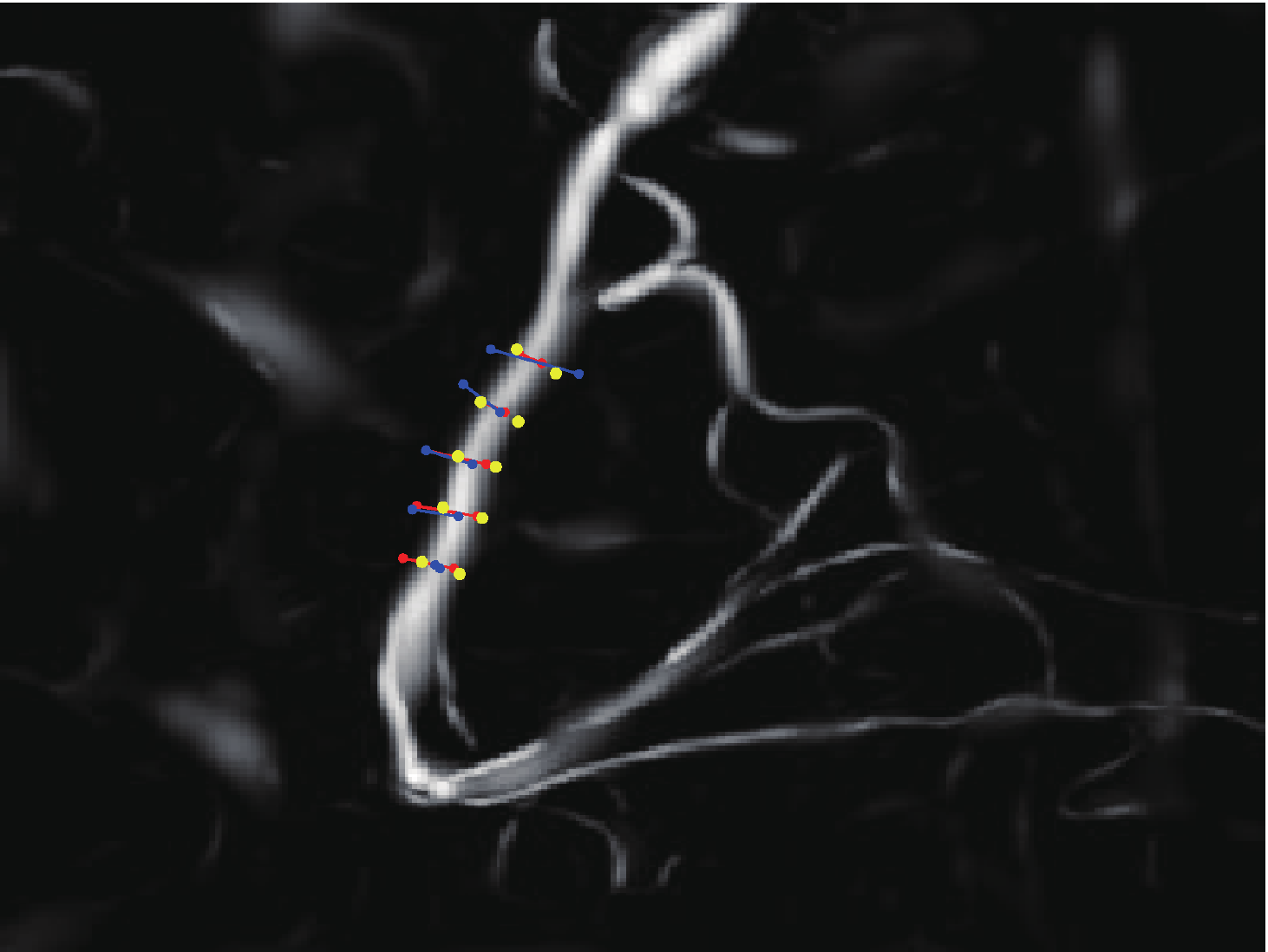} }   
	\subfigure[(a4)] { \label{fig10(a4)}     \includegraphics[width=1.15in]{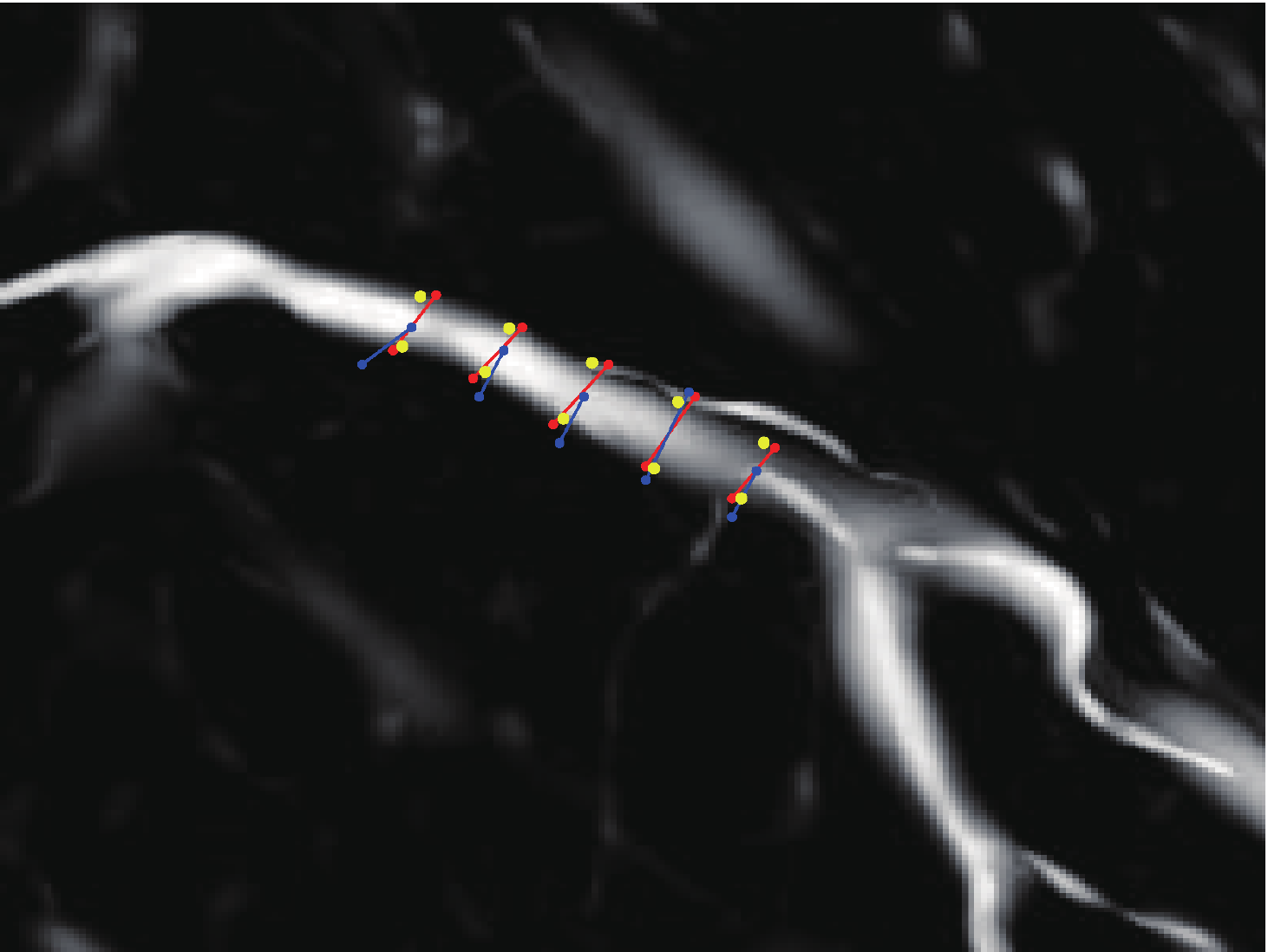} }   
	\subfigure[(b1)] { \label{fig10(b1)}     \includegraphics[width=1.15in]{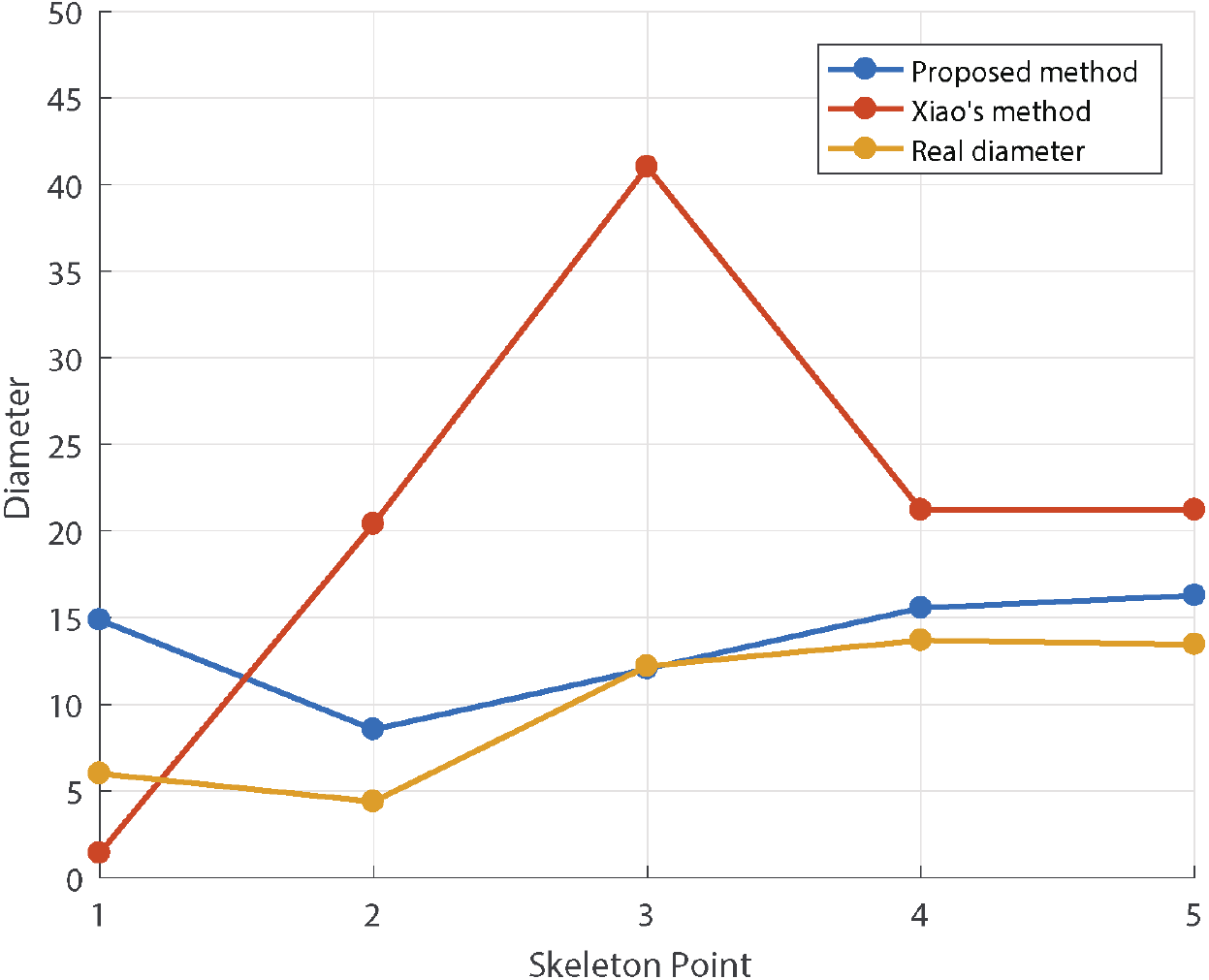} }
	\subfigure[(b2)] { \label{fig10(b2)}     \includegraphics[width=1.15in]{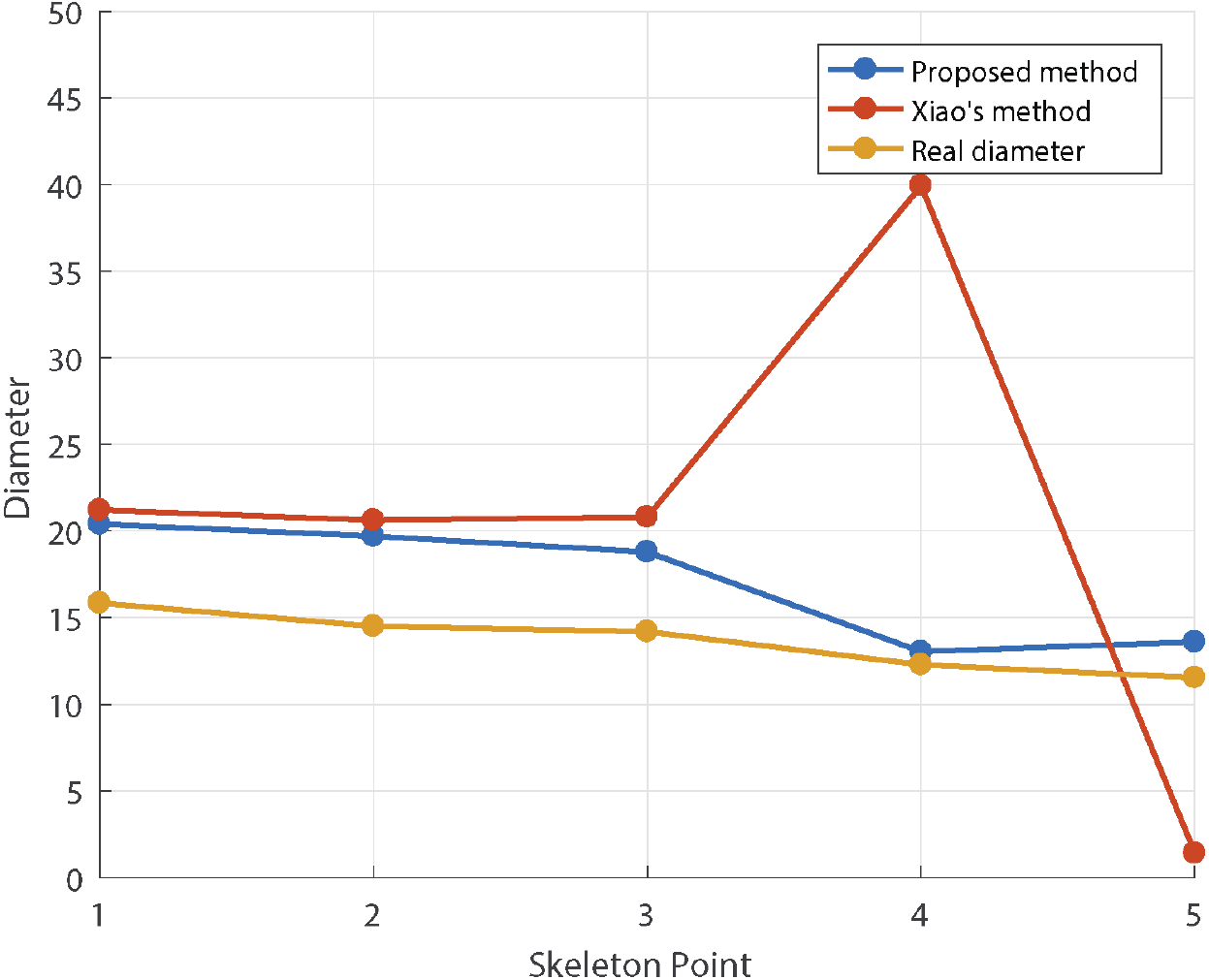} }   
	\subfigure[(b3)] { \label{fig10(b3)}     \includegraphics[width=1.15in]{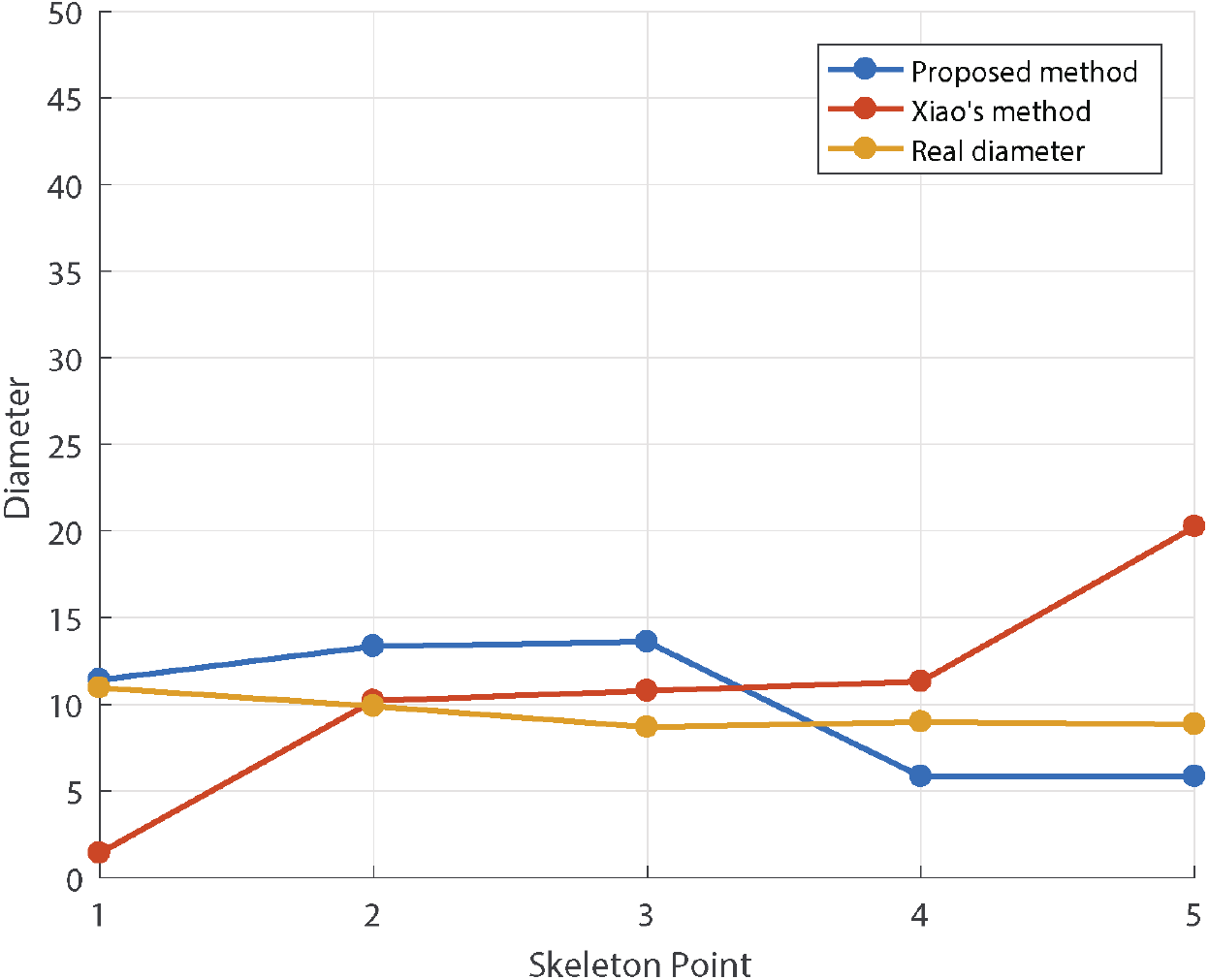} }   
	\subfigure[(b4)] { \label{fig10(b4)}     \includegraphics[width=1.15in]{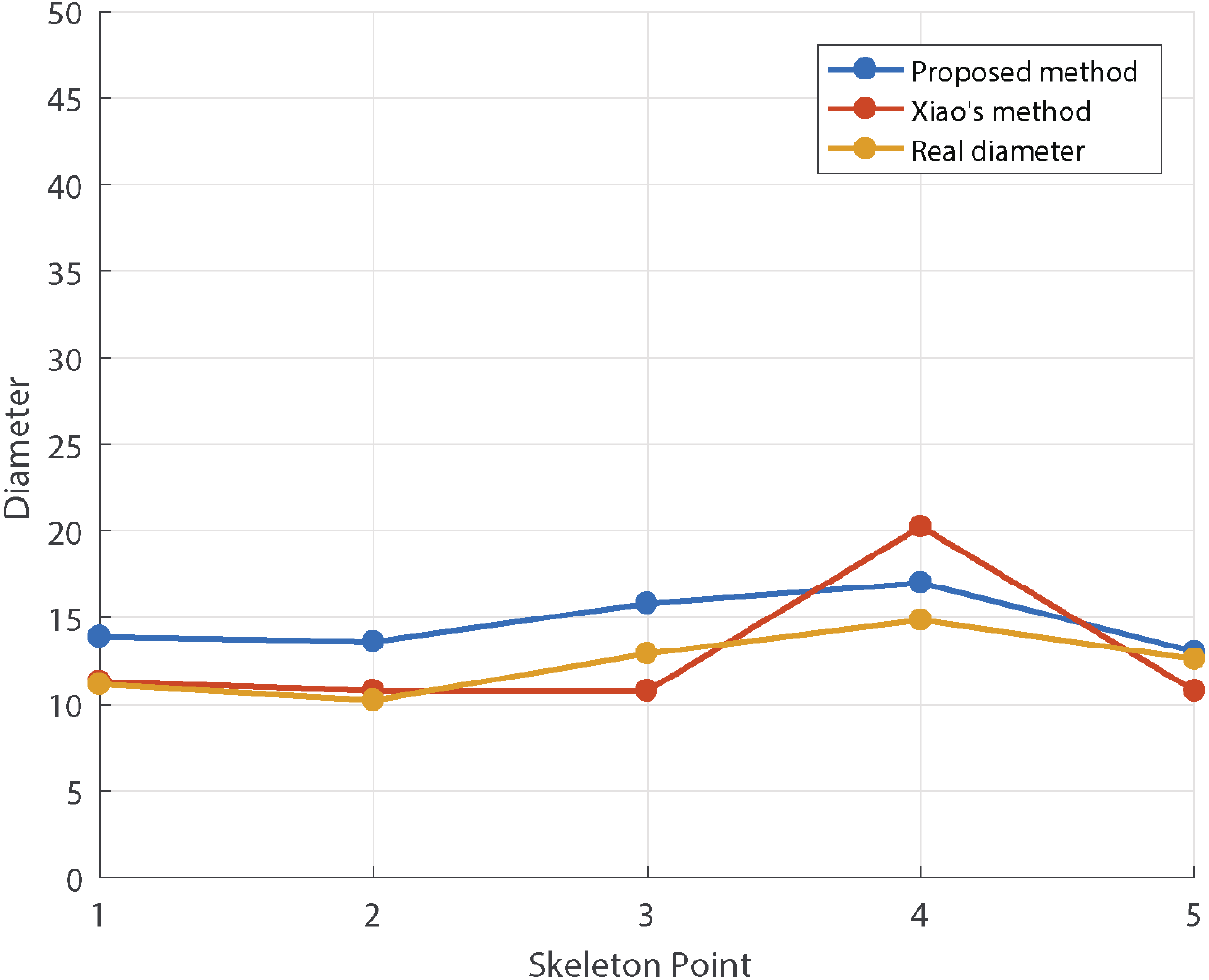} }   
	\centering
	\caption{Comparison of the methods of diameter measurement. (a) Visual results (the red, blue, and yellow dots represent the intersection points obtained by the proposed method, the points obtained by Xiao's method, and the artificially marked points, respectively). (b) Line charts of the diameters (the abscissa is the serial number of the tracking points, the unit of diameters is a pixel).}
	\label{Verification results}
\end{figure*}

\section*{Results and discussion}
\label{sec4}

\subsection*{Dataset and settings}
\label{sec4.1}

The proposed methods are evaluated on a coronary angiograms dataset consisting of 53 grayscale images sized at $400 \times 400$ pixels with available ground truth for stenosis detection
by an expert. The images are selected from CDs in DICOM format of 30 patient cases provided by Qilu Hospital (Qingdao). 

In initial  direction detection, search radius $r$ and angle $\Delta \theta$ are set to $5$ pixels and $45 ^\circ$. In bifurcation detection, $(r_1,r_2)$ and $\Delta \theta'$ are set to $(7,12)$ pixels and $135 ^\circ$ similarly. The values of thresholds     $I_0,\tau_P,\tau_1,\tau_2,d,\tau_B,\tau_3,\lambda,\tau_d$ are set to $10,4,45^\circ,30^\circ,5,2,0.8,10000,5$. All the experiments are implemented on MATLAB software of version 2019b.

\subsection*{Preliminary results}
\label{sec4.2}

To make a preliminary test on the effectiveness of our proposed solution, we selected three images with large differences in vessel structures for independent experiments and the results are shown in Fig.\;\ref{Methods results}. As can be seen from the figures, even though the vessel structures are different, the whole algorithm in this paper still has a good experimental effect on each image. Next, we will give more detailed and quantitative results of our methods.



\subsection*{Study of automatic stenosis detection}
\label{sec4.3}

To further test the effectiveness of the proposed automatic method, we adopt Xiao's method \cite{xiaoAutomaticVasculatureIdentification2013}, i.e., replace our tracking method and method of diameter measurement with hers in automatic stenosis detection, as a baseline. Fig.\;\ref{Automatic results} shows the comparison of automatic stenosis detection between the proposed method and Xiao's method. We observe that our method can detect stenoses more accurately and make fewer mistakes. This is because we have a better tracking method and more precise diameter measurement, which will be discussed later.

For quantitative evaluations, we use three metrics, i.e., sensitivity(Sen), precision(Pre), and $F_1$-score, to measure the performance of the stenosis detection on the two methods. They are expressed as follows:

\begin{equation}
	\begin{aligned}
		Sen = \frac{TP}{TP+FN} 
	\end{aligned}
\end{equation}
\begin{equation}
	\begin{aligned}
		Pre = \frac{TP}{TP+FP} 
	\end{aligned}
\end{equation}
\begin{equation}
	\begin{aligned}
		F_1 = 2 \times \frac{Pre \times Sen}{Pre+Sen}
	\end{aligned}
\end{equation}
where TP (true positive) is the number of stenoses that are detected correctly, FN (false negative) is the number of stenoses that are not detected, and FP (false positive) is the number of the points that are falsely detected as stenoses.

\begin{table}
\centering
\caption{Comparison of automatic detection methods using three metrics}
\label{Detection metrics}
\begin{tabular}{llll}
\hline\noalign{\smallskip}
Algorithm & $Sen$ & $Pre$ & $F_1$ \\
\noalign{\smallskip}\hline\noalign{\smallskip}
Xiao's Method($\tau_3=0.8$) & 0.466 & 0.697 & 0.592 \\
Proposed Method($\tau_3=0.5$) & 0.515 & 0.857 & 0.604 \\
Proposed Method($\tau_3=0.8$) & 0.757 & 0.821 & 0.788 \\
\noalign{\smallskip}\hline
\end{tabular}
\centering
\end{table}

Table.\;\ref{Detection metrics} shows that the results of the proposed method are better than Xiao's method on all metrics. And the proposed method($\tau_3=0.8$) improves the sensitivity while only sacrificing a little precision compared to ($\tau_3=0.5$). As we focus more on detecting all the stenoses, it is accepted to make some incorrect predictions.

\subsection*{Study of interactive stenosis detection}
\label{sec4.4}
To prove that the interactive method can do more precise detection than the automatic method, we use the interactive method to detect the stenoses that are not detected by the automatic method. The results are shown in Fig.\;\ref{Interactive results}. As can be seen, the interactive method can detect the labeled stenoses that are not detected by the automatic method.

The comparison of the detection results before and after the algorithm improvement, i.e., the introduction of the potential energy function, are shown in Fig.\;\ref{Comparison interactive}. By comparison, we can see that the improved algorithm has good robustness for arc structure, bifurcation structure, and even ring structure with large curvature.

To further evaluate the method of vessel diameter measurement, we compared the proposed method with Xiao's method. The artificially marked intersection points are ground truth represented by the yellow dots shown in Fig.\;\ref{Verification results}. In Fig.\;\ref{Verification results}, it can be observed that the points obtained by the proposed method are much closer to the marked ones, and so are the diameters.

To quantify the accuracy of the diameter estimation, we calculated the relative error($RE$), which is given by:
\begin{equation}
	\begin{aligned}
	RE=\frac{|R_e\left( P \right) -R_r\left( P \right) |}{R_r\left( P \right)}, 
	\end{aligned}
\end{equation}
where $R_e\left( P \right)$ and $R_r\left( P \right)$ are the estimated and real vessel diameter at point $P$. The smaller values of $ER$ indicate better measures with lower estimation errors. Experiments were carried out on twenty vessel segments randomly selected from the dataset. Table.\;\ref{Quantitative comparison} gives the quantitative results, which contain the maximum, minimum, mean, and standard deviation (std) of $RE$. We can see that the proposed method achieved lower ER values than Xiao's method on almost every vessel segment, consistent with the visual results shown in Fig.\;\ref{Verification results}. 

Overall, the results show that the proposed methods can handle more complex structures and have higher accuracy and stonger robustness than the former methods.

\begin{table}
\centering
\caption{Relative error comparison of diameter measurement methods.}
\label{Quantitative comparison}
\begin{tabular}{ccccccccc}
\hline
\multirow{2}{*}{Vessels} &
\multicolumn{4}{c}{Proposed Method} &
\multicolumn{4}{c}{Xiao's Method} \\
\cline{2-9}
& min & max & mean & std  & min & max & mean & std \\ \hline
V1 & 0.1549 & 0.8973 & 0.5209 &  0.2975 & 0.6672 & 3.0833 & 1.6888 & 1.0290 \\ 
V2 & 0.1264 & 0.6425 & 0.2720 & 0.2477 & 0.3399 & 3.2212 & 1.4208 & 1.2498 \\
V3 & 0.3021 & 0.7053 & 0.5607 &  0.1446 & 0.4446 & 3.0981 & 1.3213 & 0.9948 \\
V4 & 0.6921 & 0.7715 & 0.7217 & 0.0353 & 0.3063 & 0.8792 & 0.6210 & 0.2401 \\
V5 & 0.4498 & 0.5608 & 0.5176 & 0.0493 & 0.7539 & 4.5036 & 2.0560 & 1.7592 \\
V6 & 0.0639 & 0.1347 & 0.1082 & 0.0339 & 0.8848 & 2.7491 & 1.7200 & 0.9756 \\
V7 & 0.6102 & 0.7695 & 0.6817 & 0.0791 &  0.7755 & 1.2286 & 0.9379 & 0.200 \\
V8 & 0.2112 & 0.6163 & 0.4706 & 0.1596 & 1.8589 & 8.8833 & 4.5699 & 2.9516 \\
V9 & 0.0894 & 0.7293 &  0.3589 & 0.2469 & 0.7707 & 0.9005 & 0.8330 & 0.0515 \\
V10 & 0.0472 & 0.6809 & 0.4393 & 0.2668 & 0.7968 & 2.4050 & 1.7085 & 0.8012 \\
V11 & 0.0559 & 0.3039 & 0.2197 & 0.1018 & 0.8043 & 2.2939 & 1.4188 & 0.7808 \\
V12 & 0.0006 & 0.1345 & 0.0542 & 0.0587 & 0.5475 & 2.4841 & 1.0875 & 0.7914 \\
V13 & 0.5295 & 0.7150 & 0.6488 & 0.0733 & 0.6296 & 1.6901 & 1.0550 & 0.3971 \\
V14 & 0.0939 & 0.2676 & 0.1764 & 0.0674 & 0.6328 & 0.8850 & 0.8059 & 0.1073 \\
V15 & 0.4422 & 0.7050 & 0.5906 & 0.1094 & 0.7904 & 2.7043 & 1.2029 & 0.8405 \\
V16 & 0.5225 & 0.6906 & 0.6222 & 0.0790 & 0.7693 & 0.8777 & 0.8299 & 0.0434 \\
V17 & 0.0139 & 2.2251 & 0.4812 & 0.8641 & 0.0313 & 1.3661 & 0.7532 & 0.5476 \\
V18 & 0.1841 & 0.7091 & 0.4239 & 0.1956 & 0.8519 & 2.9889 & 1.3315 & 0.9289 \\
V19 & 0.1112 & 0.6417 & 0.4568 & 0.1823 & 0.8236 & 5.7977 & 2.1899 & 2.0836 \\
V20 & 0.0197 & 0.1703 & 0.0864 & 0.0711 & 0.2219 & 1.6625 & 1.0041 & 0.6869 \\
\cline{1-9}
Mean & 0.2360 & 0.6535 & 0.4206 & 0.1682 & 0.6851 & 2.6851 & 1.4278 & 0.8730 \\ \hline
\end{tabular}
\end{table}


\section*{Conclusions} 
\label{sec5}
Detection and quantitative analysis of stenoses are of great significance to assist clinical diagnosis of CAD. In this paper, two new stenosis detection methods were proposed. The automatic stenosis detection method can automatically extract the coronary artery tree, calculate the diameters and stenotic degrees, and mark the stenoses. With the interactive stenosis detection method, only the starting point and the ending point need to be set artificially to get the target vessel segment and detect and analyze its stenoses. A large number of experiments showed that the two algorithms have strong robustness and high accuracy for vessels with different structures.

However, the methods in this paper also have some limitations, such as high dependence on the preprocessing and the vessel contour extraction algorithm, which will produce errors under some complex vessel structures. Therefore, in the following work, we will consider using more advanced  methods like deep learning for preprocessing and vessel contour extraction to further improve the detection effect.

\noindent \textbf{Funding} Yaofang Liu, Xinyue Zhang, Wenlong Wan, Shaoyu Liu, Yingdi Liu, Hu Liu,and Xueying Zeng was supported by the National Natural Science Foundation of China [No.11771408], the Fundamental Research Funds for the Central Universities [No.201964006]. Qing Zhang  was supported by the National Natural Science Foundation of China [No.81671703], the Key Research and Development Project of Shandong Province [No.2015GSF118026], the Qingdao Key Health Discipline Development Fund, and People's Livelihood Science and Technology Project of Qingdao [No.18-6-1-62-nsh].

\noindent \textbf{Availability of data and materials} Data is not available for this study.

\section*{Declarations}
\noindent \textbf{Conflict of interest} Authors do not have any conflicts of interest.

\noindent \textbf{Authors’ contributions}
Yaofang Liu, Xinyue Zhang, and Wenlong Wan contributed equally.

\noindent \textbf{Ethical approval} This article does not contain any studies with human participants or animals performed by any of the authors.

\noindent \textbf{Informed consent} Not applicable.

\noindent \textbf{Code availability} Code is currently not made available.


\bibliographystyle{IEEEtran}
\bibliography{mybibfileWithoutHref}


\end{document}